\documentclass[conference]{IEEEtran}
\IEEEoverridecommandlockouts
\usepackage[hidelinks]{hyperref}
\usepackage{amsmath,amsfonts,amsthm}
\usepackage{paralist}
\usepackage{graphicx}
\usepackage{textcomp}
\usepackage{algorithm}
\usepackage[noend]{algpseudocode}
\usepackage{graphicx}
\usepackage{xcolor}
\usepackage{xspace}
\usepackage{tikz}
\usepackage{forest}
\usepackage{multirow}
\usepackage{verbatim}
\usepackage{balance}  
\usepackage{enumitem}
\usepackage{subcaption}

\newcommand{\spara}[1]{\noindent\textbf{{\quad}#1.}}
\newcommand{\ms}[1]{{\color{purple} [#1 - Marco]}}
\newcommand{\jl}[1]{{#1}}

\DeclareMathOperator{\EX}{\mathbb{E}}

\newtheorem{theorem}{Theorem}     
\newtheorem{lemma}[theorem]{Lemma}
\theoremstyle{definition}

\usepackage[firstpage]{draftwatermark}

\begin{document}

\SetWatermarkText{
\hspace*{6.5in}\raisebox{10.25in}{\includegraphics[scale=.25]{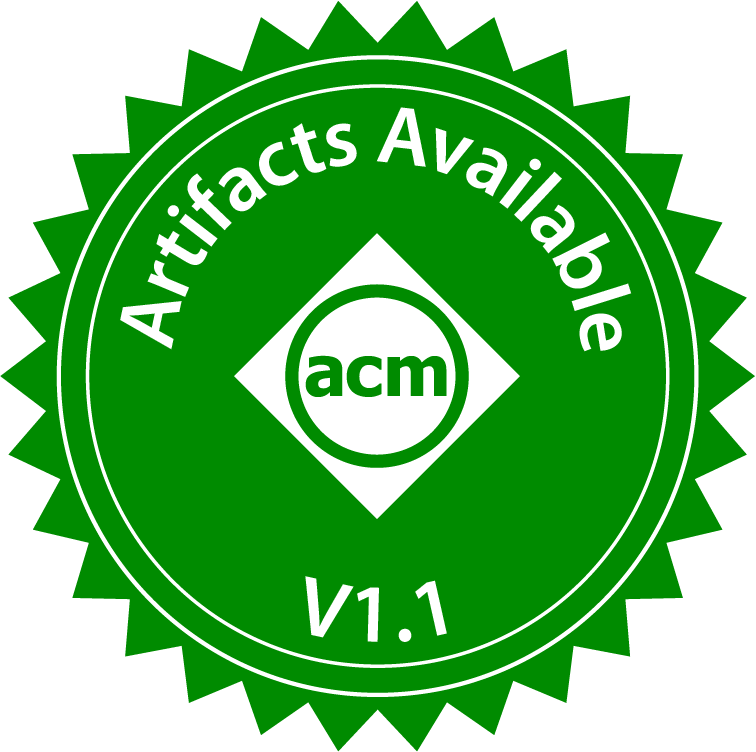}\includegraphics[scale=.25]{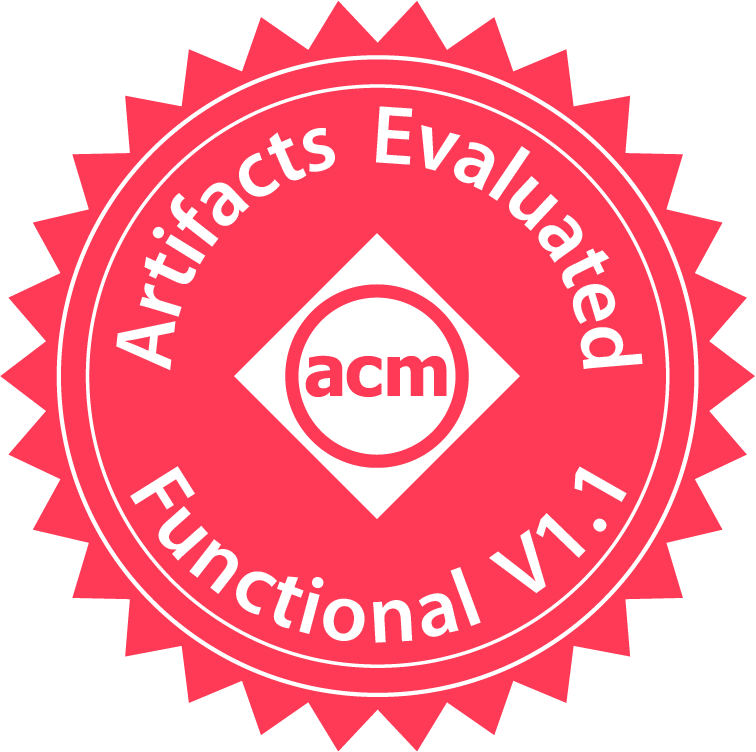}}}
\SetWatermarkAngle{0}

\title{GraphMini: Accelerating Graph Pattern Matching Using Auxiliary Graphs}

\author{\IEEEauthorblockN{Juelin Liu, Sandeep Polisetty, Hui Guan, and Marco Serafini}
\IEEEauthorblockA{\textit{Manning College of Information and Computer Sciences} \\
\textit{University of Massachusetts Amherst}\\
Amherst MA, USA \\
\{juelinliu, spolisetty, huiguan, marco\}@cs.umass.edu}
}

\maketitle

\begin{abstract}
    Graph pattern matching is a fundamental problem encountered by many common graph mining tasks and the basic building block of several graph mining systems. 
    This paper explores for the first time how to proactively prune graphs to speed up graph pattern matching by leveraging the structure of the query pattern and the input graph. 
    We propose building auxiliary graphs, which are different pruned versions of the graph, during query execution.
    This requires careful balancing between the upfront cost of building and managing auxiliary graphs and the gains of faster set operations.
    To this end, we propose GraphMini, a new system that uses query compilation and a new cost model to minimize the cost of building and maintaining auxiliary graphs and maximize gains.
    Our evaluation shows that using GraphMini can achieve one order of magnitude speedup compared to state-of-the-art subgraph enumeration systems on commonly used benchmarks.
\end{abstract}


\section{Introduction} 
\label{section:introduction}
Graph pattern matching is a fundamental problem encountered by many common graph mining tasks and real-world applications, such as analyzing social networks \cite{social1, social2} and protein-protein interaction networks~\cite{ppt1}. 
It is also the basic building block of several graph mining systems~\cite{Peregrine, AutoMine, GraphZero, Dryadic}.
The problem is challenging because the search space and the size of intermediate data increase exponentially with the size of the data graph and query graph. 

Graph pattern matching involves performing a large number of \emph{set operations} on the adjacency lists in the input graph.
These set operations are a well-known performance bottleneck and speeding them up has been the focus of prior research~\cite{SIMD}.
State-of-the-art work reduced the number of set operations by caching intermediate results to eliminate redundant operations~\cite{AutoMine} and by using better query execution schedules~\cite{GraphPi}.
Other work has proposed better implementations of the set operation~\cite{SIMD,EmptyHeaded,FlexMiner,DIMMining}.

This paper explores a different and complementary direction to speed up set operations: reducing the size of the adjacency lists that are given as input to those operations.
Graph pattern matching systems usually match one query vertex or edge at a time, using nested loops. 
Our basic insight is that, when we match a query vertex in a loop, we can already identify some data vertices in the data graph that \emph{will never be matched in deeper loops}.
We can thus \emph{proactively prune} these vertices from adjacency lists to make the adjacency lists shorter, accelerating the set operations involving these lists in deeper loops without changing the query results. 
We propose a new data structure \textit{auxiliary graph} to keep track of pruned adjacency lists on the fly during query execution. 
We materialize the idea in GraphMini, a novel single-host graph pattern-matching system that generates query execution code to efficiently build, maintain, and reuse auxiliary graphs.

GraphMini addresses several research questions.
First, we establish criteria to identify which adjacency lists can be pruned and how to prune them safely, without changing the result of a query.
The technique maintains multiple auxiliary graphs from pruning the same adjacency list, each auxiliary graph tailored to speed up the matching of a specific query vertex.
These auxiliary graphs can be further pruned online at deeper loops as more intermediate results become available. 
Our experiments show that the memory overhead of storing auxiliary graphs is only a small fraction of the data graph.

Pruning adjacency lists preserves correctness but it does not always speed up computation.
Building an auxiliary graph entails paying an upfront cost that should be amortized in deeper loops.
Deciding if the gains outweigh the costs is challenging because we cannot ascertain how often an auxiliary graph will be reused in the future.
Pruning always does not yield consistent speedups.
We thus build a cost model that leverages runtime statistics to estimate the gains of pruning an adjacency list before it is actually pruned. 
We show that using the cost model outperforms simple heuristics.

Our GraphMini system also implements a set of compile-time optimizations to minimize the computation costs in building, managing, and retrieving auxiliary graphs. 
These optimizations include removing runtime checks for retrieving the most suitable version of the pruned adjacency lists, reusing auxiliary graphs to build other auxiliary graphs, and techniques to balance load across parallel worker threads. 

Our evaluation shows that GraphMini outperforms state-of-the-art systems like GraphPi~\cite{GraphPi} and Dryadic~\cite{Dryadic} by up to 30.6x and 60.7x respectively.
In addition, the code generation algorithm has a minimal impact on the end-to-end query execution time.

To summarize, we make the following contributions: 
\begin{compactitem}
    \item We propose proactive graph pruning via a new data structure called auxiliary graph to substantially speed up graph pattern matching without changing the query results.
    \item We propose a cost model to estimate whether the cost of pruning an adjacency list will be amortized.
    \item We implement a novel single host graph pattern matching system called GraphMini, which generates query execution code tailored to maximizing the benefit of proactive pruning using several compile-time optimizations.
    \item We experimentally show that GraphMini is substantially faster than state-of-the-art approaches.  
\end{compactitem}

\section{Background and Motivation} \label{section:background}

\subsection{Problem Definition}
\label{sec:problem-def}
The graph pattern matching, or subgraph enumeration, problem takes a data graph $G$ and a query graph $q$ as input and outputs the subgraphs $\{g\}$ that are \emph{isomorphic} to $q$. 
In the \emph{edge-induced} variant, a subgraph $ g(V(g), E(g)) $ of $G$ is \emph{isomorphic} to $q$ if there exists a bijection $M$ between the sets $V(q)$ and $V(g)$ such that:
\begin{equation}
    \forall v_i, v_j \in V(q): (v_i, v_j) \in E(q) \iff (u_i, u_j) \in E(g),
\end{equation}
where $u_i = M(v_i)$ and $u_j = M(v_j)$. 

The \emph{vertex-induced} variant of the problem has the following additional requirements:
\begin{equation}
    \forall u_i, u_j \in V(g): (u_i, u_j) \in E(g) \iff (u_i, u_j) \in E(G).
    \label{eqn:vertex-centric}
\end{equation}
In this case, we say the subgraph $g$ \emph{matches} $q$. 

GraphMini supports both variants. 
Different graph mining problems require different variants~\cite{Arabesque}.
For example, the edge-induced variant is used for the frequent subgraph mining problem and the vertex-induced variant is used for motif counting.

Typically, we output a result set $R$ consisting of \emph{unique} subgraphs -- that is, subgraphs that are not \emph{automorphic} to any other subgraph in $R$.
Two subgraphs $g(V(g),E(g))$ and $g'(V(g'), E(g'))$ of $G$ are automorphic if and only if $V(g) = V(g')$ and $(u_i,u_j) \in E(g) \iff (u_i,u_j) \in E(g')$. We remove automorphic subgraphs in the output because they are essentially the same subgraph obtained by matching the query vertices in different orders. 

\subsection{Workflow of Graph Pattern Matching} 
\label{sec:workflow}
State-of-the-art graph pattern matching and graph mining systems~\cite{AutoMine, GraphZero, Dryadic, GraphPi} typically run in two stages: \textit{query scheduling} and \textit{query execution}. 
GraphMini also follows these two stages with the additional step of creating and using auxiliary graphs to accelerate query execution. 

\spara{Query scheduling} 
Given an arbitrary query graph $q$ as an input, query scheduling aims to find a \textit{matching order} for the query. 
A matching order specifies which vertex in the query graph should be matched first to a vertex in the data graph.
Given a matching order $\phi$, we use the notation $v_i$ to denote the query vertex in $V(q)$ that has the $i^{th}$ position in the matching order.
If the query graph $q$ is symmetrical, there could be automorphic matches.
Query scheduling will generate a set of \emph{canonicality checks}, which are rules to prune symmetrical matches. We call the matching order of query vertices and the canonicality checks a \emph{query schedule}. 

\spara{Query execution} 
Query execution first generates a query execution plan specific to a query schedule and then executes the generated plan to produce matches. 
The state-of-the-art graph pattern matching systems \cite{AutoMine, GraphZero, GraphPi} generate a query execution plan as a sequence of nested loops using Ullmann's backtracking algorithm \cite{Ullmann}. The algorithm materializes a depth-first algorithm that traverses vertices in the data graph and tries to identify a match between the vertices in the data graph and those in the query graph. 

\begin{figure}[h]
    \centering
    \includegraphics[width=\linewidth]{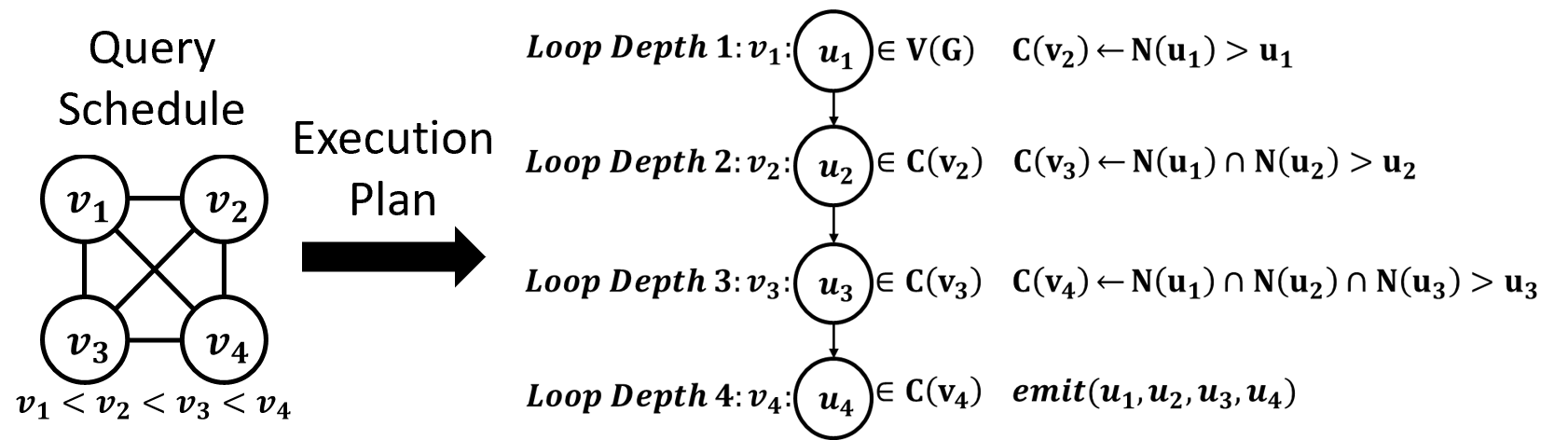}
    \captionof{figure}{Query Schedule and Execution Plan a 4-clique Query}
    \label{figure:4clique-plan}
\end{figure}
\spara{Example}
Figure~\ref{figure:4clique-plan} shows an example of a query schedule and execution for the 4clique pattern. 
All vertices in the query pattern are assigned a sequence number $v_i$.
The canonicality constraint $v_i < v_{i+1}$ means that for each pair of vertices $u_i$, $u_{i+1}$ in the data graph that match $v_i$ and $v_{i+1}$ respectively, the vertex id of $u_i$ is smaller than the id of $u_{i+1}$.
This ensures that only one canonical permutation of the match $[u_1, \ldots, u_4]$ will be returned as a result, while all its other automorphic permutations will be filtered out.

During query execution, the first outer loop iterates over all $u_1$ vertices in the data graph. 
Each of them matches the first query vertex $v_1$.
Each vertex in $N(u_1)$, which is the adjacency list of $u_1$, is a candidate for matching the second query vertex $v_2$ if it respects the canonicality constraint $u_1 < u_2$.
We use $C(v_2)$ to denote the \emph{candidate set} for $v_2$.
The second loop iterates over each $u_2 \in C(v_2)$ and performs similar steps.


\subsection{Challenges and Opportunities}
We now explain the challenges of graph pattern matching and the opportunities that motivate the design of the GraphMini system.

\subsubsection{Challenges}
Set operations represent the main bottleneck of query execution.
State-of-the-art systems~\cite{GraphPi, AutoMine, GraphZero, Dryadic} use \emph{prefix sets} to cache the results of set operations and reuse them at lower loops. 
For example, at loop depth $2$ in Figure~\ref{figure:4clique-plan} it is possible to cache the result of the set operation $N(u_1) \cap N(u_2)$ and reuse it in the lower loop to avoid executing the same operation again.

We find that, even with existing prefix set-based optimizations, \emph{accessing adjacency lists, rather than prefix sets, still dominates computational cost}.
Table~\ref{tab:vertex_scanned} shows the fraction of vertices accessed from adjacency lists vs. prefix sets by set operations while running clique counting queries on various graphs using Dryadic~\cite{Dryadic}. 
The results show that scanning vertices from adjacency lists corresponds to a large fraction (43\%-95\%) of the total vertices scanned.
Therefore, pruning adjacency lists has the potential to speed up the execution of set operations.

\begin{table}[h]
\centering
\tabcolsep=0.08cm
\small 
\begin{tabular}{|ccccccc|}
\hline
\multicolumn{7}{|c|}{\textbf{Percentage of Vertices Scanned from Adjacency Lists}} \\ \hline
\multicolumn{1}{|c|}{\textbf{Graph}} &
  \multicolumn{1}{c|}{\textit{Wiki}} &
  \multicolumn{1}{c|}{\textit{Patents}} &
  \multicolumn{1}{c|}{\textit{YouTube}} &
  \multicolumn{1}{c|}{\textit{Lj}} &
  \multicolumn{1}{c|}{\textit{Orkut}} &
  \textit{Friendster} \\ \hline
\multicolumn{1}{|c|}{\textit{4-Clique}} &
  \multicolumn{1}{c|}{81\%} &
  \multicolumn{1}{c|}{43\%} &
  \multicolumn{1}{c|}{86\%} &
  \multicolumn{1}{c|}{91\%} &
  \multicolumn{1}{c|}{89\%} &
  82\% \\ \hline
\multicolumn{1}{|c|}{\textit{5-Clique}} &
  \multicolumn{1}{c|}{91\%} &
  \multicolumn{1}{c|}{45\%} &
  \multicolumn{1}{c|}{90\%} &
  \multicolumn{1}{c|}{95\%} &
  \multicolumn{1}{c|}{95\%} &
  90\% \\ \hline
\end{tabular}
\caption{Cost of set operations: fraction of vertices accessed from adjacency lists vs. prefix sets.}
\label{tab:vertex_scanned}
\end{table}

\subsubsection{Opportunities}

Motivated by the above observation, the basic idea of GraphMini is to prune adjacency lists at the upper loop levels to accelerate set operations involving these lists in deeper loop levels. 
Even though we may not know exactly which vertices match a given query vertex beforehand, we can often identify some vertices that \emph{cannot} match it. 

To see why, consider the example of Figure~\ref{figure:4clique-plan}.
The first loop of the query will iterate over all vertices matching the query vertex $v_1$.
In the Figure, $u_1$ is an alias denoting any data vertex that matches $v_1$.
Suppose that we match the data vertex $a = u_1$ to the query vertex $v_1$ in an iteration of the first loop.
After that, the algorithm will execute the nested loops to find all subgraphs matching the query where $v_1$ is matched to $a$.
Then, the algorithm will backtrack to the first loop and iterate over to the next match for $v_1$.

After matching $v_1$ to $a$, we can already conclude from the structure of the query graph that any vertex that matches $v_4$ will have to be also in $N(a)$.
This is because $v_1$ and $v_4$ are neighbors in the query graph.
In other words, the candidate set $C(v_4)$ of data vertices that match the query vertex $v_4$ must be a subset of $N(a)$.
Our goal is to use this information to prune the adjacency lists that will be used to compute $C(v_4)$. 

The candidate set $C(v_4)$ is computed at loop depth 3 by intersecting the set $N(u_1) \cap N(u_2)$, which is computed at loop depth 2, with $N(u_3)$.  
To speed up this set intersection, we can compute at loop depth $1$ a pruned version of $N(u_3)$ that only includes candidates in $C(v_4) \subseteq N(a)$ as:
$$P(u_3) = N(u_3) \cap N(a).$$
We can then use this shorter pruned adjacency list instead of $N(u_3)$ to compute $C(v_4)$.

This example shows that adjacency lists can be pruned \emph{online}, during query execution, and \emph{proactively}, before the adjacency lists are used.
In Section~\ref{sec:pruning}, we show how to generalize this simple example to arbitrary patterns and graphs.
In the example, we want to prune the adjacency list of $N(u_3)$ at loop depth 1 but we will only know which vertices will be matched to $u_3$ in the future, when executing the nested loop at depth 3.
We show how to find a set of candidate adjacency lists to prune at loop depth 1.
We also show how to leverage prefix sets to prune adjacency lists more effectively.

\spara{Weighing benefits and costs}
It is not straightforward to determine whether the upfront cost to compute and store the pruned list will be amortized in the future.
In the previous example, the cost of computing and storing the pruned adjacency list $P(u_3)$ at loop depth 1 can be later amortized only if pruning significantly reduces the size of $N(u_3)$, and the pruned adjacency list $P(u_3)$ will be used multiple times to compute $C(v_4)$ in lower loops.

In Section~\ref{sec:cost-model}, we propose a cost model to estimate the gain of building a pruned adjacency list.
The cost model is used online to decide whether to prune an adjacency list. 
In the example, we use the cost model at loop 1 to decide whether to prune $N(u_3)$ or not.

\spara{Managing multiple auxiliary graphs}
Managing pruned adjacency lists could potentially add considerable runtime complexity.
One source of complexity is dealing with multiple pruned versions of the same adjacency list.
Suppose that we have computed $P(u_3)$ in the first loop, moved to the second loop, and matched a vertex $u_2$.
We can further incrementally prune $P(u_3)$ by intersecting it with $N(u_2)$.
Therefore, we need to keep multiple versions of the same pruned adjacency list at different loop depths and we need to retrieve them efficiently to support fast incremental pruning.
Another source of complexity is that we need different pruned adjacency lists for different query vertices.
In the example, the pruned adjacency list $P(u_3)$ can only be used to find candidates of the query vertex $v_4$ after matching $v_1$ with $a$.

Keeping track of all these versions and using the right version at each set operation could potentially add a significant runtime cost.
In Section~\ref{sec:reduceoverhead}, we discuss how the code generator of GraphMini avoids these costs by determining which version to use at compile time.
We also discuss how it uses nested loops to effectively balance work among worker threads.

\section{Overview of GraphMini}
GraphMini is an efficient graph pattern matching system that leverages auxiliary data structures to accelerate query execution. 
The fundamental idea of GraphMini is to build auxiliary graphs consisting of pruned adjacency lists and use them in deeper nested loops to reduce execution time. 
This approach does not remove edges from the original data graph. 
Instead, it builds multiple versions of pruned graphs to accelerate the computations for different prefix sets and candidate sets in the query. 

\begin{figure}[h]
    \centering
    \includegraphics[width=\linewidth]{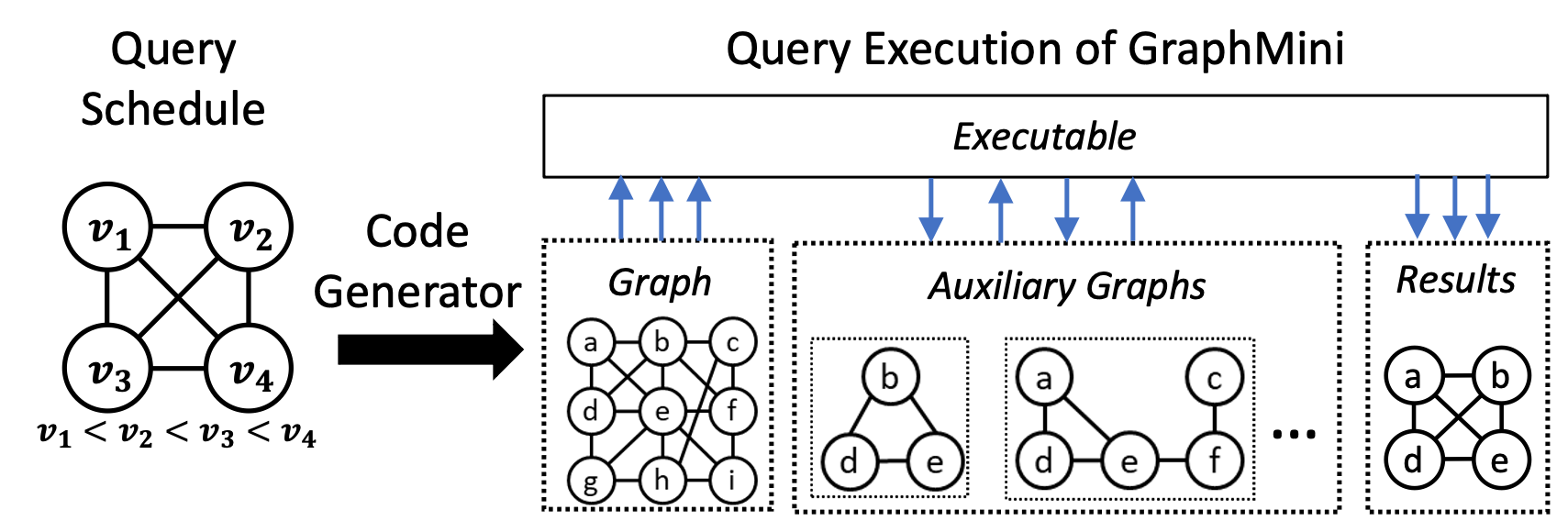}
    \captionof{figure}{Workflow of GraphMini}
    \label{figure:workflow}
\end{figure}

The overview of GraphMini is shown in Figure~\ref{figure:workflow}. 
It takes as input a query pattern graph and a data graph and produces a set of matched subgraphs. 
Our contribution lies in the \textit{code generator}, which produces an efficient subgraph enumeration algorithm using auxiliary graphs. 
The \emph{code generator} produces an efficient executable for each input query pattern. 
GraphMini then uses the generated executable to match the input query on the data graph. 
During query execution, GraphMini manages multiple versions of auxiliary graphs to accelerate the set operations without changing the results. 

\section{Proactive Online Pruning}
\label{sec:pruning}

\begin{table}[t]
\centering
\tabcolsep=0.05cm
\resizebox{\columnwidth}{!}{%
\begin{tabular}{|l|l|}
\hline
\textbf{Notation}                       & \textbf{Description}                                            \\ \hline
$q(V(q), E(q))$                          & a query graph, its vertices and its edges                            \\ \hline
$G(V(G), E(G))$                          & data graph, its vertices and its edges                       \\ \hline
$M(v)$                                   & data vertex matching the query vertex $v$                          \\ \hline
$A_h(v_k, v_i)$                          & auxiliary graph with selecting set $C_h(v_k)$ and filtering set $C_h(v_i)$           \\ \hline
$h, k, i$                                  & three indices such that $h < k < i$                              \\ \hline
$v_i$                                    & $i^{th}$ query vertex to be matched ($v_i \in V(q)$)             \\ \hline
$u_i$                                    & a data vertex matched to $v_i$ ($u_i \in V(G)$)                       \\ \hline
$I_i$                                    & a partial match $[u_1, u_2, \ldots, u_i]$               \\ \hline
$C(v_i)$                                 & candidate set of $v_i$                                           \\ \hline
$C_h(v_i)$                               & the $h^{th}$ prefix-set for $v_i$                      \\ \hline
$N(u_i)$                                 & neighbors of the vertex $u_i$                                    \\ \hline
$P_h(u_k|v_i)$                             & pruned neighbors of $u_k$ at depth h for constructing $C_k(v_i)$ \\ \hline
\end{tabular}%
}
\caption{Notations}
\label{notations}
\end{table}

This section defines auxiliary graphs and discusses how to find all potential adjacency lists to prune.
Table~\ref{notations} summarizes the notations used in the following discussions.
 
\subsection{Basic Pattern Matching}
We first introduce how state-of-the-art systems perform graph pattern matching~\cite{AutoMine,Dryadic} with prefix set-based optimizations.
Auxiliary graphs build on top of these optimizations.

\spara{Computing Candidate Sets}
State-of-the-art systems match a query vertex at a time and compute its candidate set using nested loops.
Consider computing the candidate set $C(v_i)$ at the nested loop at depth $i-1$.
Assume the algorithm has already found a partial match $I_{i-1}= [u_1, \ldots, u_{i-1}]$.
The set $C(v_i)$ is computed by performing set operations on the adjacency lists of the vertices in $I_{i-1}$.
The code generator analyzes the edges between $v_i$ and the vertices in the query graph that have already been matched, $[v_1, \ldots v_{i-1}]$.

Specifically, the algorithm partitions the vertices of $I_{i-1}$ into two sets $I_T$ and $I_F$ such that:
\begin{gather}
    \forall u \in I_T: u = M(v_h) \text{ and } (v_h, v_i) \in E(q) \notag\\
    \forall u \in I_F: u = M(v_h) \text{ and } (v_h, v_i) \notin E(q)  \label{equation:partition-partial-matching-candidate}
\end{gather}

In the \emph{vertex-induced} variant, the candidate set $C(v_i)$ contains the data vertices matching the query vertex $v_i$ and is defined as:
\begin{equation}
    \label{equation:func_1}
    C(v_i) = V(G) \cap \big( \bigcap_{u \in I_T} N(u) \big) \setminus \big( \bigcup_{u \in I_F} N(u) \big). 
\end{equation}

In the \emph{edge-induced} variant, the candidate set $C(v_i)$ is defined as:
\begin{equation}
    \label{equation:func_2}
    C(v_i) = V(G) \cap \big( \bigcap_{u \in I_T} N(u) \big) \big). 
\end{equation}

Na\"{i}vely calculating candidate sets based on Eqns.~\ref{equation:func_1} and~\ref{equation:func_2} is slow. 
We now discuss techniques and optimizations proposed by the existing work. 

\spara{Connected Ordering} \label{connected_ordering}
In real graphs, the size of $V(G)$ is usually orders of magnitude larger than the largest adjacency list in the graph. 
To avoid using $V(G)$ as input in the set operation, schedulers put a constraint on the ordering.
Each matched vertex $v_i$ must be a neighbor of some matched $v_h$ where $h<i$ unless it is the top vertex, that is, $i=1$. 
This constraint ensures that $I_T$ contains at least one vertex, so the computation for $C(v_i)$ becomes: 

\begin{equation}
    C(v_i) =  \bigcap_{u \in I_T} N(u) \:\setminus \bigcup_{u \in I_F} N(u). \label{eqn:naive_candidate_computation}
\end{equation}

\spara{Prefix sets} 
\label{section:prefix-optimization}
Existing systems~\cite{AutoMine, GraphPi, GraphZero, Dryadic} use prefix sets to avoid the redundant execution of set operations. 
It pushes the set operations in computing candidate sets to the uppermost loop where the computation is feasible and reuses the intermediate results in the inner loops.

We use $C_{h}(v_i)$ to denote the prefix set for the candidate set $C(v_i)$ that can be computed at the nested loop with depth $h$, with $h < i$. $C_{h}(v_i)$ is the superset of $C(v_i)$ that can be obtained based on the adjacency lists of the vertices that have been matched at loop depth $h$.  $I_h$ can be partitioned into two sets $I_{h, T}$ and $I_{h, F}$ such that: 

\begin{gather}
    \forall u \in I_{h, T}: u = M(v_j) \text{ and } (v_i, v_j) \in E(q), \notag\\
    \forall u \in I_{h, F}: u = M(v_j) \text{ and } (v_i, v_j) \notin E(q).  \label{equation:partition-partial-matching-prefix}
\end{gather}

The prefix set $C_h(v_i)$ is only materialized when $I_{h,T} \neq \emptyset$. It is defined as:
\begin{equation}\label{eqn:naive_prefix_computation}
    C_h(v_i) = \bigcap_{u \in I_{h,T}} N(u) \: \setminus \bigcup_{u \in I_{h,F}} N(u). 
\end{equation}
When $h = i - 1 $, the prefix set $C_h(v_i)$ is the candidate set $C(v_i)$.
The condition for the materialization of a prefix set is that $I_{h, T} \neq \emptyset$.
This ensures that the prefix set can be used to actually prune the candidate set.

Instead of computing the prefix set $C_h(v_i)$ from scratch at each depth $h$, we can compute it incrementally based on the prefix set at the depth $h-1$ and the adjacency list of the data vertex $u_h$: 
\begin{equation}
C_h(v_i) = C_{h-1}(v_i) \circ_h N(u_h),
\label{eqn:prefix_prefix_computation}
\end{equation}
In vertex-induced graph pattern matching, the $\circ_h$ operator is the set intersection operator ($\cap$) if $(v_i, v_h) \in E(q)$ or the set subtraction operator ($\setminus$) otherwise.
In the edge-induced variant, if $(v_i, v_h) \notin E(q)$ there is no set subtraction.

\subsection{Auxiliary Graphs} 
\label{sec:auxiliarygraph}
An \emph{auxiliary graph} is a data structure designed to accelerate set operations by reducing the size of the adjacency lists they take as input. 
Each auxiliary graph consists of multiple \emph{pruned adjacency lists}, each storing only a subset of the neighbors of a data vertex.
Whenever possible, GraphMini uses auxiliary graphs instead of the original graph to find the adjacency lists for the set operations.
Pruned adjacency lists are smaller than the original lists in the graph so set operations are faster.

Auxiliary graphs and prefix sets are fundamentally different and complementary optimizations.
Auxiliary graphs speed up set operations by pruning their inputs; prefix sets execute set operations as early as possible in the loop hierarchy.
Auxiliary graphs can speed up the calculation of prefix sets. 

\spara{Definition of Auxiliary Graph}
Each auxiliary graph is relative to a \emph{pair} of query vertices, which we denote as $v_i$ and $v_k$ in the following.
The auxiliary graph $A(v_k, v_i)$ contains the pruned adjacency lists of some data vertices $u_k$ matching $v_k$.
These pruned lists are then used instead of $N(u_k)$ to compute prefix sets or candidate sets for $v_i$.
The way the pruning occurs depends on how $v_i$ and $v_k$ are connected in the query graph.

The auxiliary graph optimization has two main steps. 
At each loop depth $h$, we compute some pruned adjacency lists. 
Then, at each loop depth $k > h$, we use those pruned adjacency lists to speed up the computation of that set. 
The basic idea poses two core questions: (i) which adjacency lists to prune, and (ii) how to prune the adjacency lists to ensure correctness. We next answer these two questions and discuss our solutions. 

\spara{Identifying Potential Adjacency Lists to Prune}
\label{sec:eager}
We now show how to find all \emph{potential} adjacency lists to prune. 
We discuss how to decide whether we should \emph{actually} prune them in Section~\ref{sec:cost-model}.

Consider building an auxiliary graph $A_h(v_k,v_i)$ at some loop depth $h$.
Our goal is to prune away some of the vertices in $N(u_k)$, where $u_k$ matches $v_k$, in order to accelerate the computation of $C_k(v_i)$ at a deeper loop depth $k > h$. Specifically, we would like prefix set calculation to become: 
\begin{align}
    &\text{At loop h:} \quad \text{Compute } P_{h,k}(u_k|v_i) \subset N(u_k), \\ 
    &\text{At loop k:} \quad C_k(v_i) = C_{k-1}(v_i) \circ_k P_{h,k}(u_k|v_i).
    \label{eqn:target}
\end{align}
As we will see in the following, we prune $N(u_k)$ in a way that is specific to the query vertex $v_i$ at loop $h$.
We then want to replace the adjacency list $N(u_k)$ with a pruned version $P_{h,k}(u_k|v_i)$ in Eqn.~\ref{eqn:prefix_prefix_computation} at loop $k$ to reduce the size of the input of the set operation $\circ_k$ and thus its running time.

To prune $N(u_k)$, we need to know the vertex $u_k$.
At loop $h$, however, we only have a partial match $I_h =[u_1 \ldots u_h]$ which is a prefix of $I_k$ and does not include $u_k$.
The question is, how can we identify $u_k$ so that we can prune its adjacency list?

Our proposed solution is to use $C_h(v_k)$ to identify the set of vertices that $u_k$ can be matched to. In other words, we can potentially prune all $\{N(u) \: | \: u \in C_h(v_k)\}$. The solution is based on the properties of prefix sets.
The data vertex $u_k$ is defined to be one of the vertices matching the query vertex $v_k$.
It holds that:
\begin{equation}
u_k \in C(v_k) = C_{k-1}(v_k) \subseteq C_h(v_k).
\label{eqn:identify-lists-to-prune}
\end{equation} 
When using the prefix optimization, the candidate set $C(v_k)$ is equal to the last of its prefix sets $C_{k-1}(v_k)$.
The prefix sets are built to incrementally remove candidate vertices using set operations, so $C(v_k) \subseteq C_h(v_k)$.

\spara{Pruning the Adjacency Lists}\label{sec:pruning_the_adjacency_lists}
After identifying the potential adjacency lists to prune, the next question is how to prune them. 
Even though at loop $h$ we do not know exactly which vertices match $v_i$, we can already tell the vertices that are not in $C_h(v_i)$ are definitely not matches, so we can remove them from $N(u_k)$.
The result will not change because of the following observation from the set theory. 


\begin{lemma}[Superset-based pruning rule]
\label{lem:pruning}
    Let $C$, $C'$, $P$, and $N$ be four sets, and $\circ$ denote either the set intersection or the set subtraction operator.
    If $C \subseteq C'$ and $P = C' \cap N$ then: 
    $$
        C \circ N = C \circ P
    $$
\end{lemma}

We use this rule to compute a pruned version of $N(u_k)$ for the auxiliary graph of $v_i$ at loop $h$.
We call this the \emph{pruned adjacency list} $P_{h,k}(u|v_i)$ and define it as follows:
\begin{equation}
    P_{h,k}(u|v_i) = C_h(v_i) \cap N(u), \quad u\in C_h(v_k).
    \label{eqn:prune-compute}
\end{equation}
Note that we use the notation $u$ instead of $u_k$ because $u$ is in the prefix set of $v_k$ at loop $h$ but it may not end up being a candidate for $v_k$.
Each pruned adjacency is specific to a query vertex $v_i$ because it is filtered using a prefix set of $C(v_i)$.

At loop $h$, we compute the pruned adjacency list.
Later, we use the pruning rule again at loop $k$ and replace Eqn.~\ref{eqn:prefix_prefix_computation} with the following expression:
\begin{equation}
    C_k(v_i) = C_{k-1}(v_i) \circ_k P_{h,k}(u_k|v_i)
    \label{eqn:prune-use}
\end{equation}
This expression is equivalent to Eqn.~\ref{eqn:prefix_prefix_computation} because of the superset pruning rule (Lemma~\ref{lem:pruning}) since $ C_k(v_i) \subseteq C_{k-1}(v_i)$.

In summary, we still compute $ C_k(v_i)$ incrementally from $C_{k-1}(v_i)$ as in Eqn.~\ref{eqn:prefix_prefix_computation}, but we now do it using the pruned adjacency list $P_{h,k}(u_k|v_i)$ to speed up the execution of the set operation $\circ_k$.

\spara{Putting It All Together}
The auxiliary graph $A_h(v_k,v_i)$ at loop depth $h$ consists of the pruned adjacency lists $P_{h,k}(u|v_i)$ for some $u\in C_h(k)$.
It optimizes the set operation $\circ_k$ in two steps:
\begin{enumerate}
    \item At loop $h$, we prune $N(u)$ by computing $P_{h,k}(u|v_i)$ as in Eqn.~\ref{eqn:prune-compute}, for each $u \in C_h(v_k)$. 
    \item At loop $k > h$, we compute $C_k(v_i)$ using $P_{h,k}(u|v_i)$ as in Eqn.~\ref{eqn:prune-use}.
\end{enumerate}

Building an auxiliary graph requires two prefix sets, which we call the \emph{selecting set} and the \emph{filtering set}.
The selecting set is the prefix set $C_h(v_k)$ in Eqn.~\ref{eqn:identify-lists-to-prune} used to select which adjacency lists to prune. 
The filtering set is the prefix set $C_h(v_i)$ in Eqn.~\ref{eqn:prune-compute}, which is used to compute the pruned adjacency lists before adding them to the auxiliary graph. \jl{Both the \emph{selecting set} and the \emph{filtering set} must be materialized before we can build an auxiliary graph.}

\begin{figure}[h]
    \centering
    \includegraphics[width=0.8\linewidth]{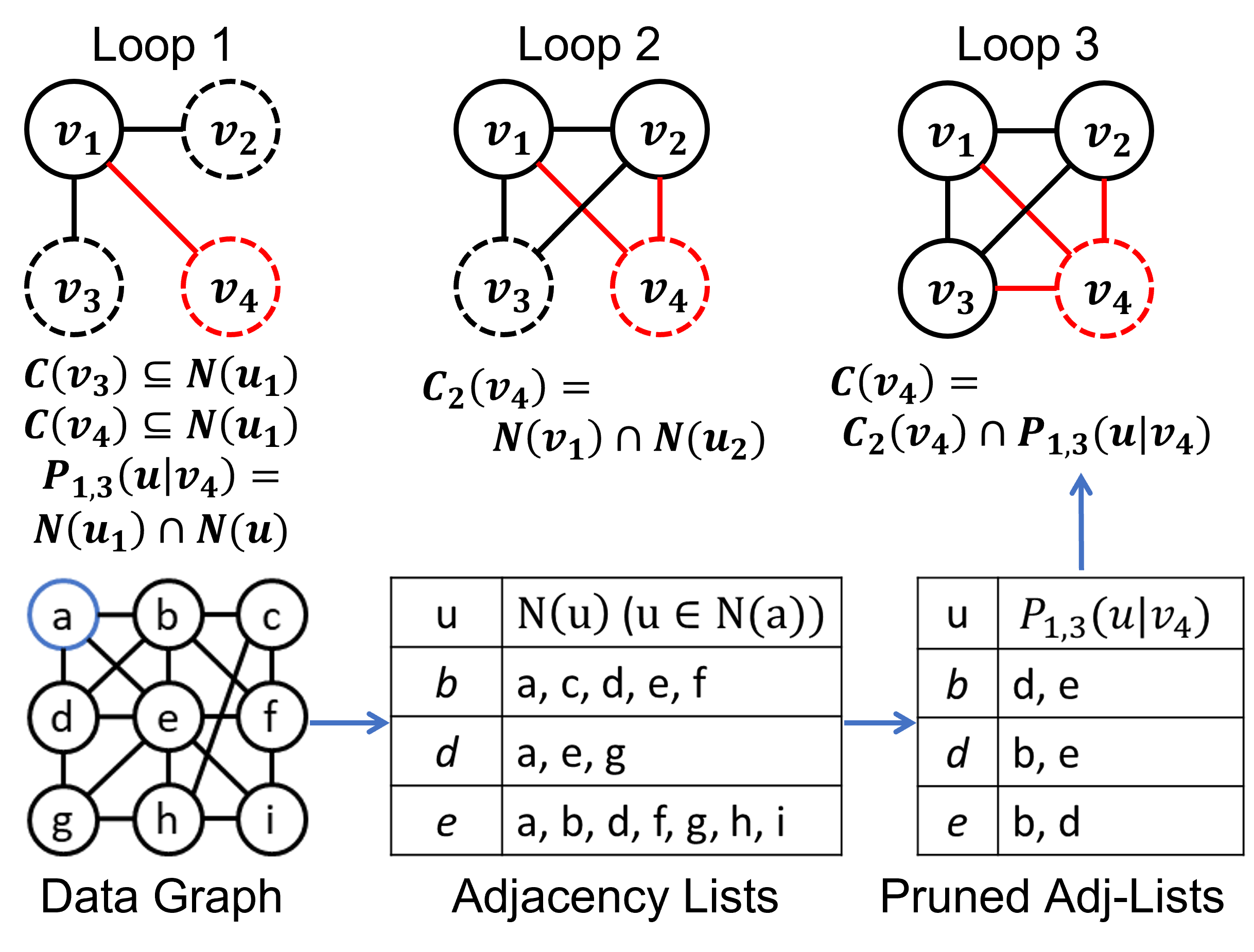}
    \captionof{figure}{GraphMini Pruning Example. Query vertices with dashed outlines have not been pruned in that loop. We focus on the set operations for computing $C(v_4)$ at different loop levels. We assume $u_1$ is matched to $a$ in this example.}
    \label{figure:example}
\end{figure}

\spara{Example} 
We use the example in Figure~\ref{figure:example} to illustrate how the proposed auxiliary graph works. For simplicity, we ignore canonicality checks in this example. 
We assume $u_1=a$, $C(v_3)$ and $C(v_4)$ will always be a subset of $N(a)$ in the inner loops because the query schedule requires that $u_3$ and $u_4$ must be the neighbors of $u_1$. 
To accelerate the computation for $C(v_4)$, we can prune the adjacency lists of each vertex in $N(a)$, which are $N(b)$, $N(d)$, and $N(e)$.
We do that by intersecting each of them with $N(a)$ and obtain pruned adjacency lists $P_{1,3}(u|v_4)$, with $u \in \{b,d,e\}$. 
Then, we can use these pruned lists to replace $N(u_3)$ in the computation for the candidate set $C(v_4)$ without affecting the results. Figure~\ref {figure:example} shows the original adjacency lists and their pruned version.
\section{Pruning Cost Model}
\label{sec:cost-model}
In the previous Section, we have discussed how to find the set of adjacency lists to \emph{potentially} prune.
In this section, we discuss how to choose adjacency lists to actually prune based on our cost-benefit analysis.
We analyze the cost and benefit of pruning an adjacency list. 
Based on the analysis, we propose an \emph{online} approach to decide which adjacency lists to prune using runtime information. 

\subsection{Benefit and Cost of Pruning}


Pruning an adjacency list requires a set intersection, whose cost is a function of the size of its inputs. 
The benefits depend on how many vertices we can prune from the adjacency list and on how many times the pruned adjacency list will be reused. 
We express costs and benefits in terms of the reduction in the number of comparisons required to perform this computation. 
We assume that the input adjacency lists and prefix sets are sorted and we estimate the number of vertices to be scanned by the set operations as the sum of the set sizes.

\spara{Definitions}
When our algorithm computes a pruned adjacency list at loop $h$, it reaches that loop depth after obtaining a partial match $I_h$.
The pruned adjacency list will be used in the nested loops at a higher depth than $h$, where all the partial and final matches will be extensions of $I_h$. 
We use $e(I_h, k, u)$ to denote the set of all partial matches of size $k$ that extend $I_h$ where the data vertex $u$ matches the query vertex $v_k$.
Note that each prefix set $C_j(v_i)$ is calculated after having obtained a partial match $I_j$, for any value of $j$.
We say in that case that $C_j(v_i)$ is relative to $I_j$. 
We still use the notation $C_j(v_i)$ instead of a more precise $C_{j, I_j}(v_i)$ for simplicity.

\spara{Cost/Benefit Analysis}
We start by counting the number of comparisons performed at loop depth $k$ without auxiliary graphs, using the expression of Eqn.~\ref{eqn:prefix_prefix_computation}.

\begin{equation*}
    o(I_h,k,u,i) = \sum_{I_k \in e(I_h,k,u)} |C_{k-1}(v_i)| + |N(u)|
\end{equation*}

We assume that the sets $C_{k-1}(v_i)$ and $N(u)$ are sorted lists and that the set operation scans them.
The worst-case running time of the set operation is thus the sum of the cardinality of the two sets.

With auxiliary graphs, the cost becomes:
\begin{align*}
    a(I_h,k,u,i) =& \: |C_h(v_i)|+|N(u)| +\\
    & \sum_{I_k \in e(I_h,k,u)} |C_{k-1}(v_i)|+|C_h(v_i) \cap N(u)|
\end{align*}
The first line of the equation is the cost of computing the pruned adjacency lists at loop $h$ using Eqn.~\ref{eqn:prune-compute}.
The second line is the cost of using the pruned adjacency list, which has size $|C_h(v_i) \cap N(u)|$.
For each vertex $u_k \in C(v_k)$, we compute $C_k(v_i)$ using Eqn.~\ref{eqn:prune-use}.

The gain of building a pruned adjacency list at loop $h$ can thus be expressed as follows.
\begin{align*}
    g(I_h,k,u,i)  =& \: o(I_h,k,u,i) - a(I_h,k,u,i)\\
                =& \: |e(I_h, k, u)| \cdot (|N(u)| - |C_h(v_i)\cap N(u)|)\\
                -& \: (|C_h(v_i)|+|N(u)|).
\end{align*}

This expression shows that pruning is not always advantageous. Pruning all $u \in C_h(v_k)$ does not guarantee that there is a gain.
Ideally, we would like to compute $P_{h,k}(u|v_i)$ if and only if the gain is positive for a given vertex $u$. 

 \subsection{Online Cost Model}\label{onlineApproach}
 We now discuss how we use runtime information to predict the gain of a pruning adjacency list. Our intuition is to use runtime information to estimate the variables in the above analysis to compute the gain.
 
 The above analysis needs four variables to estimate the gain: $|e(I_h, k, u)|,\: |N(u)|,\:|C_h(v_i)\cap N(u)|$ and $ |C_h(v_i)|$. At runtime, we can directly obtain $|N(u)|$ and $|C_h(v_i)|$ by reading from the data graph and the prefix set. 
 However, $|C_h(v_i) \cap N(u)|$ and $|e(I_h, k, u)|$ are unknown unless we perform further computation to obtain those sets, which can be costly. 
 We now discuss how we estimate the sizes of these two sets at loop $h-1$ using information that is already available.

Assuming that each vertex in $N(u)$ is equally likely to be in $C_h(v_i)$, we can estimate $|C_h(v_i) \cap N(u)|$ as:
\begin{equation*}
    |C_h(v_i) \cap N(u)|  \approx \frac{|C_h(v_i)| \cdot |N(u)|}{|V(G)|}
\end{equation*}

To estimate $|e(I_h, k, u)|$, we assume that all the vertices in $C_h(v_k)$ are equally likely to appear in $C(v_k)$ as we extend $I_h$ to $I_k$. 
However, the cardinalities of the candidate sets $C(u_{h+2}),\dots, C(u_k)$ are unknown at the time we run the cost model since those candidate sets will only be materialized in the inner loops. 
We estimate the size of these candidate sets using their prefix sets at loop $I_h$ as follows: 

\begin{align*}
    |e(I_h, k, u)|  \approx  \frac{1}{|C_h(v_k)|} \times \prod_{i\in[h+1,k]} \EX{(|C(v_i)|)} 
\end{align*}
where $\EX{(|C(v_i)|)}$ denotes the estimated cardinality of the candidate set $C(v_i)$.

We estimate the cardinality of a candidate set $\EX{(|C(v_i)|)}$ as follows.  
First, we compute some query-independent statistics of the data graph offline~\cite{GraphPi}.
Assuming the degree of adjacency lists have a uniform distribution, let $p_1$ be the probability that two randomly selected vertices are connected and $p_2$ be the probability that two randomly selected vertices are connected given that they are both connected to a common vertex (which forms a triangle). We have:

$$p_1 = \frac{2 \times |E(G)|}{|V(G)|^2},\quad p_2 = \frac{tri\_cnt \times |V(G)|}{ (2 \times |E(G)|)^2}$$

\jl{The first equation calculates $p_1$ as the fraction of vertex pairs that are connected by an edge. 
It divides the total number of undirected edges in the graph by the number of unique vertex pairs that can be selected from the graph, disregarding the order of the vertices in the pair.
The second equation calculates $p_2$. 
If two vertices $x$ and $y$ have a common neighbor $z$ and are connected with each other, it implies that a triangle consisting of $x$, $y$, and $z$ exists. 
Thus, we can compute $p_2$ as the fraction of triangles including $z$ over the number of distinct pairs of neighbors of $z$.
}

Then, at query execution time, we use Algorithm~\ref{estimate-size} to estimate the sizes of the candidate sets $\EX{(|C(v_i)|)}$ in the deeper nested loops.
The algorithm uses the statistics of the data graph $p_1$ and $p_2$ as constants to estimate the pruning power of future set intersections in the inner loops. 

The algorithm estimates the size of the prefix sets of $C(v_i)$ at all nested loop depths. 
At loop depth $h$, the prefix set $C_h(v_i)$ is to be intersected with $N(u_{h+1})$ to compute $C_{h+1}(v_i)$.
We can estimate the size of the intersection result by computing the probability that a randomly selected vertex from $|C_{h}(v_i)|$ is connected to $u_{h+1}$. If $u_{h+1}$ and $u_i$ do not have a common neighbor, then:
\begin{align*}
    |C_{h+1}(v_i)| &\approx |C_h(v_i)| \times \EX{(|N(u_{h+1})|)} \div |V(G)| \\
    &= |C_h(v_i)| \times 2 \times |E(G)| \div |V(G)|^2 \\
    & = |C_{h}(v_i)| \times p_1
\end{align*}

\jl{If we know from the query graph that $u_{h+1}$ and $u_i$ must have a common neighbor $u_j \: (j \leq h)$, then we know that $u_{h+1}$ and all vertices in $|C_{h}(v_i)|$ must be connected to $u_j$. In this case, we can use $p_2$ to estimate the probability that a randomly selected vertex from $|C_{h}(v_i)|$ is connected to $u_{h+1}$:
\begin{equation*}
    |C_{h+1}(v_i)| \approx |C_{h}(v_i)| \times p_2
\end{equation*}
A similar logic is followed for all subsequent loop depths after $h+1$ until we obtain the final estimation $\EX(|C(v_i)|)$.}

The runtime overhead of this estimation is small because the algorithm multiplies $|C_h(v_j)|$ to a constant factor that is computed at compile time based on the statistics of the data graph ($p_1$ and $p_2$) and the structure of the query.

A limitation of the above analysis is it underestimates the reuse frequency for high-degree adjacency lists since both $p_1$ and $p_2$ are computed based on the assumption that the degrees are uniformly distributed.
To address the limitation, we pre-allocate the memory buffer for storing the pruned adjacency lists of high-degree vertices, but delay the actual pruning to the query time to avoid over-pruning. 


\begin{algorithm}[h]
  \caption{Candidate Set Cardinality Estimation}
  \label{estimate-size}
  \begin{flushleft}
    \textbf{Input: } $E(q)$: edges in the query graph. \\
    \textbf{Input: } $C_h(v_i)$: prefix set for candidate set $C(v_i)$. \\
    \textbf{Output: } $\EX(|C(v_i)|)$: Estimated cardinality of $C(v_i)$.
  \end{flushleft}
  
  \begin{algorithmic}
    \Procedure{EstimateCard}{$ E(q), \: C_h(v_i)$}
      \State \emph{EstCard} $\gets |C_h(v_i)|$
      \For{$l \in [h+1, i]$}
      \If{$(v_i, v_l) \in E(q)$}
        \If{$\exists v_t: (v_t, v_l), (v_t, v_i) \in E(q)$ for $t < i, l$}

        \State \emph{EstCard} $\gets$ \emph{EstCard} $\times \: p_2$
        \Else
        \State \emph{EstCard} $\gets$ \emph{EstCard} $\times \: p_1$
       \EndIf
      \EndIf
      \EndFor
      \State $\EX(|C(v_i)|) \gets$ \emph{EstCard}
      \State \Return $\EX(|C(v_i)|)$
    \EndProcedure
  \end{algorithmic}
\end{algorithm}
\section{Code Generation}
\label{sec:reduceoverhead}
We now discuss the code generation process and compile-time optimizations in the GraphMini system.
\subsection{Identifying Auxiliary Graphs}
\begin{algorithm}[t]
  \caption{Find Auxiliary Graphs}
  \label{find-auxiliary-graphs}
    \textbf{Input: } \emph{Schedule}: Query pattern schedule. \\
    \textbf{Input: } $|V(q)|$: Query pattern size. \\
    \textbf{Output: } \emph{AuxGraphs}: Candidate auxiliary graphs.
  \begin{algorithmic}
    \Procedure{FindAuxGraphs}{{\emph{Schedule}$, \: |V(q)|$}}
      \State \emph{AuxGraphs} $\gets []$
      \For{$k \in [3,|V(q)|-1]$}
        \State \emph{PrefixSets} $\gets$ all prefixes materialized at loop $k$
        \For{$C_k(v_i) \in$ \emph{PrefixSets}}
          \For{$h \in [1, k-2]$}
            \If{$C_h(v_i)$ and $C_h(v_k)$ are materialized}
                \State \emph{AuxGraphs}.add$(A_h(v_k, v_i))$
            \EndIf
          \EndFor
        \EndFor
      \EndFor
      \State \Return \emph{AuxGraphs}
    \EndProcedure
  \end{algorithmic}
\end{algorithm}

\jl{Given and query and its schedule, GraphMini generates query-specific code to find matches (see Figure~\ref{figure:workflow}).}
GraphMini runs Algorithm~\ref{find-auxiliary-graphs} at code generation time to identify candidate auxiliary graphs that can be potentially used.
It then adds static references to the auxiliary graphs to avoid the runtime cost of looking up which ones can be used.
\jl{The auxiliary graph can contain either a reference to the original adjacency list or the pruned adjacency, depending on the decisions made at runtime using the cost model discussed in Section~\ref{onlineApproach}.}

Given a query schedule, the algorithm iterates through the prefix sets that are computed at loop level $k>=3$. We skip the prefix sets at the first two loops because their inputs cannot be further pruned. For a prefix set $C_k(v_i)$ computed at loop depth $k$, we check whether we can prune the adjacency lists $N(u_k)$ at loop $h \leq k - 2$. \jl{This is done by checking if the selecting set $C_h(v_k)$ and the filtering set $C_h(v_i)$ are both materialized, as we need to use them to compute the pruned version of $N(u_k)$ (see Section~\ref{sec:auxiliarygraph}). } 
If so, we add the associated auxiliary graph to a list. Later, the code generator will insert all the auxiliary graphs in the list into the generated algorithm.

\spara{Time Complexity Analysis} The number of prefix sets for a query graph $q$ is in $O(|V(q)|^2)$. For each prefix set, we need to check if its parent prefix set and selecting set are both materialized at some upper loops. Since there are at most $|V(q)|$ loops and checking whether a prefix set is materialized can be done in $O(1)$, the time complexity of finding all the auxiliary graphs is in $O(|V(q)|^3)$. 

\subsection{Compile-Time Optimizations}
\spara{Reusing Auxiliary Graphs}
In addition to using auxiliary graphs to speed up set operations for computing prefix sets, GraphMini determines at code generation time whether it can re-use one auxiliary graph to speed up the construction of another auxiliary graph.

For example, assume we can build two auxiliary graphs at loop level $h$ and $h+1$ respectively, both auxiliary graphs intend to accelerate the computation for $C_k(v_i)$. We refer to the auxiliary graphs as $A_h(v_k, v_i)$ and $A_{h+1}(v_k,v_i)$ respectively. The key idea is that we can use $A_h(v_k, v_i)$ to speed up the construction of $A_{h+1}(v_k,v_i)$. 

Consider the auxiliary graph $A_{h+1}(v_k,v_i)$. We can utilize the pruned adjacency list $P_{h, k}(u | v_i)$ to compute the pruned adjacency list $P_{h+1, k}(u | v_i)$ since:
\begin{align*}
P_{h+1,k}(u | v_i) &=  C_{h+1}(v_i) \cap N(u) \\
    &= C_{h+1}(v_i) \cap C_h(v_i) \cap N(u) \\
    &= C_{h+1}(v_i) \cap P_{h,k}(u | v_i).
\end{align*}
Thus we can use the pruned adjacency lists in $P_{h,k}(u | v_i) \in A_h(v_k,v_i)$ to construct $A_{h+1}(v_k,v_i)$.

In summary, if both the selecting set and the filtering set of an auxiliary graph $A$ are supersets of those of another auxiliary graph $A'$, GraphMini generates code that uses $A$ to build $A'$. 

\spara{Nested Parallelism} 
\jl{Graph pattern matching algorithms execute queries using nested “for” loops as described in Section \ref{section:background}. Each loop matches one query vertex.} Existing code-generation-based graph pattern matching systems only parallelize the first loop in their generated code for the single host subgraph enumeration workload. In this case, each thread obtains one root vertex at a time and extends all its partial matches. This approach cannot effectively use all the cores in a multiple-core system when the load across different root vertices is highly skewed, as it often happens in graphs. 

Thus, our code generator uses nested loop parallelism by default to reduce the imbalance between worker threads. 
\jl{We implemented nested parallelism in two steps: during the compilation phase and then at runtime.

During the compilation phase, we track the necessary variables for executing each loop and create function object classes to encapsulate that information. For example, for a 4-clique query, we can create a function object class solely responsible for performing computations within the second, third, and fourth loops respectively. Each instance of the object will retain the necessary information, including the value of $u_1$, the prefix sets, and auxiliary graphs that were created during the execution of the upper loops. 

During runtime, an evaluation is performed to assess the number of vertices to be scanned in the second loop. When the vertex count surpasses a predefined threshold, we encapsulate the following loops within a function object instance and execute them concurrently using multiple threads. 
In the 4-clique query example, by concurrently executing the function object generated in the first step, we enable the simultaneous extension of $u_1$ into 4-cliques across multiple threads instead of using a single thread as in previous work. 
}

To reduce the overhead of obtaining buffer memory inside nested loops, we assign each worker a private memory pool to manage its allocated memories, allowing the worker to efficiently reuse its allocated buffer without synchronization with other workers. 

\jl{Note that nested parallelism and auxiliary graphs are two orthogonal techniques. 
Nested parallelism reduces workload imbalance across different threads, whereas auxiliary graphs reduce the total amount of computations.}

\section{Evaluation} 
\label{section:evaluation}
\subsection{Experiment Settings}
\spara{Compared Systems}
We compare GraphMini with two state-of-the-art systems: Dryadic~\cite{Dryadic} and GraphPi~\cite{GraphPi}. 

Dryadic~\cite{Dryadic} is a compilation-based subgraph enumeration system. It supports both vertex-induced and edge-induced subgraph matching on labeled and unlabeled graphs. We obtain the binary of Dryadic from its author. The version we obtained also comes with a backend for set operations, which implements set intersections and set subtractions as linear scans of sorted adjacency lists. 

GraphPi~\cite{GraphPi} is the state-of-the-art subgraph enumeration system. It supports edge-induced subgraph enumeration on unlabeled graphs. It also optimizes for counting if the user only wants to know the number of matches. We use the open-sourced GraphPi implementation in our evaluation. 

Our evaluation shows that when the query schedules generated by the two systems are the same (e.g. cliques), Dryadic has better performance due to its more efficient backend implementation. However, when the systems generate different schedules, GraphPi usually outperforms Dryadic. Thus, we use GraphPi's scheduling algorithm in GraphMini to generate query schedules.



\spara{Hardware and Software Setup}
We run all experiments on an AWS c6i.16xlarge instance, which is running Ubuntu 22.04 with 32 cores (64 threads) and 128 GB of RAM. 
The machine has one Intel(R) Xeon(R) Silver 8375C CPU (2.90GHz, Turbo 3.50Ghz).
All tests use 64 CPU threads unless otherwise stated. 
All software is written in C++ and compiled with GCC 11.3 with -O3 optimization. 

\spara{Multi-threaded Settings} \label{sec:impl_nested}
For all baseline systems for comparison, we use OpenMP (v4.5) to automatically parallelize the generated algorithms on the first loop \jl{by using the ``pragma omp parallel" construct.} We use the ``dynamic" scheduling strategy with a chunk size of 1 in all experiments, which provides the best performance in our setting.

For GraphMini, we use the Intel oneTBB (2021.8.0) library to implement the nested loop parallelism. For partitioning the root vertices, we use its ``simple\_partitioner" with a chunk size of 1, which is similar to the ``dynamic" scheduling in OpenMP. For deeper loop levels, we use ``auto\_partitioner" to distribute tasks to each worker. The ``auto\_partitioner" will choose the chunk size at runtime based on the computation resources available on the machine.

\spara{Data and Query Graphs}
We use six real-world data graphs from SNAP \cite{SNAP_datasets} in our evaluation. Their statistics are shown in Table \ref{graph-info}. 
Data and auxiliary graphs are stored in the main memory during query execution. 
We treat all the graphs as undirected. 

\begin{figure}[t]
    \centering
    \includegraphics[width=\linewidth]{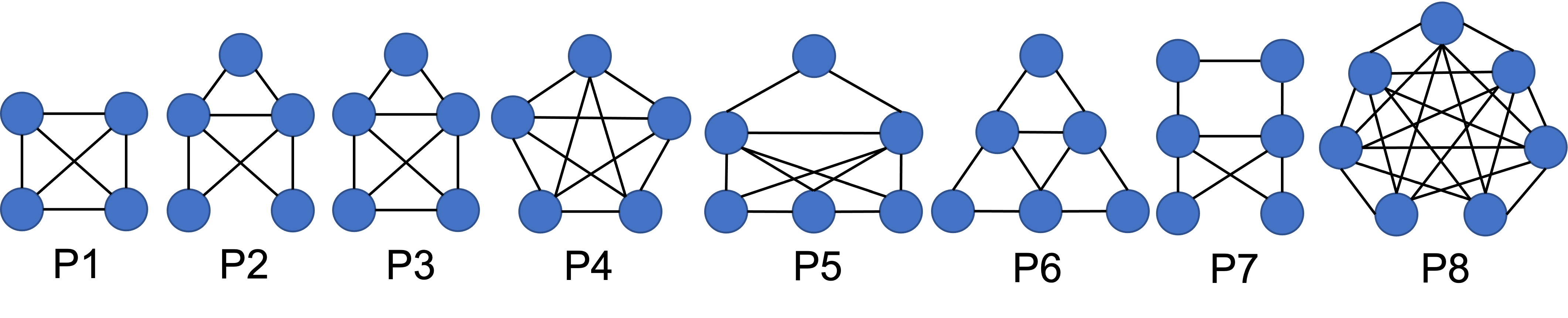}
    \captionof{figure}{Query Graph Patterns}
    \label{figure:patterns}
\end{figure}

The query graphs we use in the evaluation are shown in Figure \ref{figure:patterns}.
They come from the benchmarks used in the evaluation of previous work~\cite{GraphPi, GraphZero, Dryadic}.
They are complex patterns meant to stress test the subgraph enumeration algorithms.

\begin{table}[]
\centering
\small 
\tabcolsep=0.05cm
\caption{Data Graphs.}
\label{graph-info}
\resizebox{\columnwidth}{!}{%
\begin{tabular}{|llllll|}
\hline
Name       & $|V(G)|$ & $|E(G)|$ & $d_{max}$  &$d_{avg}$    & Description                 \\ \hline
Wiki       & 7.1k     & 103k     & 1.1k       &29    & Wikipedia votes             \\
YouTube    & 1.1M     & 3.0M     & 28.7k      &5.5    & YouTube social network      \\
Patents    & 3.8M     & 16.5M    & 0.8k       &8.7    & US Patents citation network \\
Lj         & 4.0M     & 34.7M    & 14.8k      &17    & LiveJournal social network  \\
Orkut      & 3.1M     & 117M     & 33.3k      &75    & Orkut social network        \\
Friendster & 65.6M    & 1.8B     & 5.2k      &55    & Friendster gaming network   \\
\hline
\end{tabular}%
}
\end{table}

\subsection{GraphMini vs State-of-the-Arts}
\begin{figure}
\begin{subfigure}{\linewidth}
  \centering
  \includegraphics[width=0.8 \linewidth]{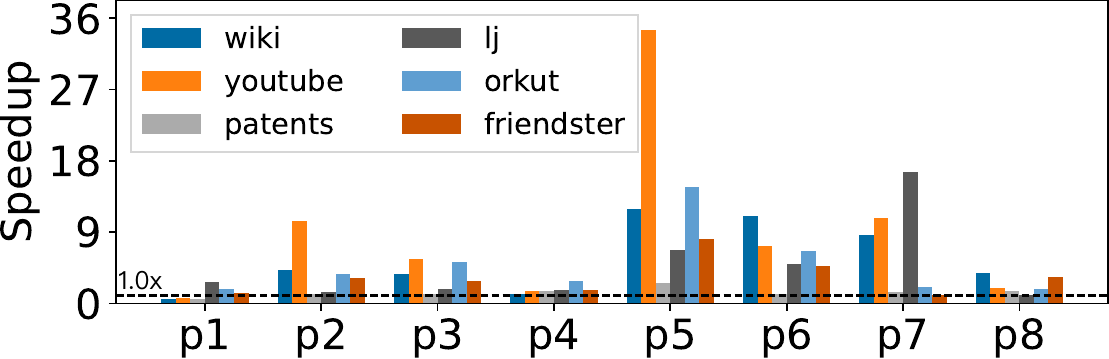}
  \caption{GraphMini vs Dryadic (Vertex-Induced)}
  \label{fig:mg_dryadic_edge}
\end{subfigure} \newline
\begin{subfigure}{\linewidth}
  \centering
  \includegraphics[width=0.8 \linewidth]{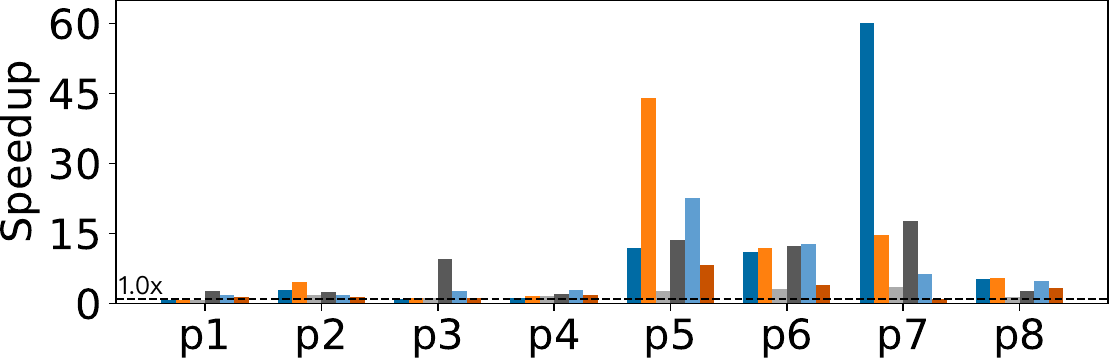}
  \caption{GraphMini vs Dryadic (Edge-Induced)}
  \label{fig:mg_dryadic_edge}
\end{subfigure} \newline
\begin{subfigure}{\linewidth}
  \centering
  \includegraphics[width=0.8 \linewidth]{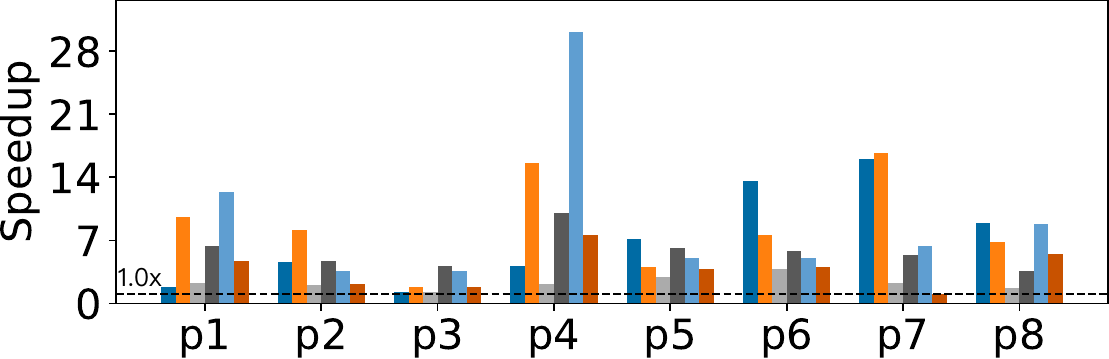}
  \caption{GraphMini vs GraphPi (Edge-Induced)}
  \label{fig:mg_graphpi_edge}
\end{subfigure} \newline
\begin{subfigure}{\linewidth}
  \centering
  \includegraphics[width=0.8 \linewidth]{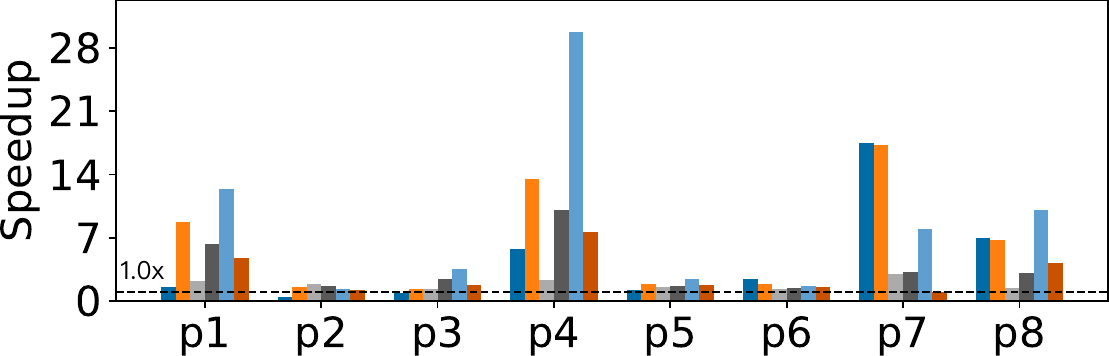}
  \caption{GraphMini vs GraphPi (Edge-Induced Counting)}
  \label{fig:mg_graphpi_edge}
\end{subfigure} 
\caption{GraphMini vs State-of-The-Art Systems}
\label{fig:mg_state_of_the_arts}
\end{figure}
\spara{Query Execution Time}
We start by comparing the time different systems take to match the query patterns of Figure~\ref{figure:patterns}, considering different variants of the pattern matching problem. GraphPi does not support vertex-induced pattern matching but it introduces dedicated optimizations for edge-induced counting, so we consider that variant separately. \jl{The execution time encompasses the time required to identify all matching subgraphs within the graph after the data graph has been loaded into memory. For GraphMini, the execution time includes all the steps required to create and maintain auxiliary graphs, which occur at runtime. }

Figure \ref{fig:mg_state_of_the_arts} shows the speedup of GraphMini when compared to the state-of-the-art systems. 
We exclude the time to load the input graph into memory to show the direct effects of the auxiliary graphs on accelerating code efficiency. For GraphMini, the reported query execution times always include the construction time of auxiliary graphs, which takes place \emph{online} during query execution. 
An empty bar in the diagram indicates that the baseline system does not finish that query in 24 hours. 

GraphMini is significantly faster than both state-of-the-art systems.
In edge-induced pattern matching, GraphMini outperforms GraphPi and Dryadic by up to 30.6x and 60.7x respectively. In vertex-induced pattern matching, GraphMini outperforms Dryadic by up to 35x. 

GraphPi uses a single runtime to execute all queries instead of generating a dedicated binary for each query. 
This approach reduces code generation overhead (see Table~\ref{tab:code_gen} for detail) but increases runtime overhead. 
For example, running $P_4$ on $Orkut$ GraphPi requires 220s to finish whereas Dryadic and GraphMini only take 19s and 6.8s respectively. 

Dryadic uses code generation but it does not leverage auxiliary graphs, so its set operations are more expensive than in GraphMini.
It also suffers from an imbalanced workload across different threads. For instance, when running vertex-induced query $P_2$ on $YouTube$, Dryadic requires 14s to finish, but the average finishing time per thread is only 2.2s, which indicates a great amount of workload imbalance. 
GraphMini only takes 1.6s using the same schedule thanks to the combined acceleration of nested parallelism and auxiliary graphs.

\spara{Code Generation Time}
\begin{table}[t]
\centering
\tabcolsep=0.08cm
\begin{tabular}{|c|c|c|c|c|c|c|c|c|}
\hline
\textit{\textbf{Pattern}} & \textit{$P_1$} & \textit{$P_2$} & \textit{$P_3$} & \textit{$P_4$} & \textit{$P_5$} & \textit{$P_6$} & \textit{$P_7$} & \textit{$P_8$} \\ \hline
\textit{$T_{GraphPi}$}   & 0.001s & 0.001s & 0.002s & 0.02s & 0.02s & 0.03s & 0.08s & 0.8s \\ \hline
\textit{$T_{Dryadic}$}   & 1.3s   & 1.4s   & 1.3s   & 1.3s  & 1.3s  & 1.5s  & 1.4s  & 1.5s \\ \hline
\textit{$T_{Online+N}$} & 1.0s   & 1.6s   & 1.7s   & 1.7s  & 1.9s  & 1.9s  & 2.0s  & 1.8s \\ \hline
\end{tabular}%
\caption{Code Generation Time}
\label{tab:code_gen}
\end{table}
During code generation, GraphMini introduces extra overhead for code generation compared to the state-of-the-art because it needs to consider auxiliary graphs.
Using nested loop parallelism also creates more code for compilation. 
Table~\ref{tab:code_gen} shows GraphMini's code generation time compared to the state-of-the-art. $T_{GraphPi}$ reports the code generation time of GraphPi~\cite{GraphPi}. $T_{Dryadic}$ reports the code generation time of Dryadic~\cite{Dryadic} the results are the average of 3 runs. 

GraphPi has the least code generation overhead because it does not need to compile source code into binaries for different queries. It only needs to generate a query schedule that can be executed directly by its backend. 
Dryadic and GraphMini generate a dedicated binary for a given query so the code generation overhead is higher when compared to GraphPi. GraphMini needs to additionally generate code for building and managing auxiliary graphs and function objects to implement nested loop parallelism, hence the code generation overhead increases as the pattern size increases.
However, the overhead of code generation can be amortized over time by reusing the generated binaries. It is also small when compared to the pattern-matching workload on large graphs. 
Running small pattern queries on small data graphs can make code generation become a bottleneck, making GraphPi's approach more attractive.

\subsection{Ablation Study}
\spara{Auxiliary Graphs and Nested Parallelism}
We now evaluate the effect of two optimizations we presented in this paper: auxiliary graphs and nested parallelism. 
We compare GraphMini with a baseline \emph{Base} that uses the same codebase but does not use those two optimizations.

GraphMini outperforms \emph{Base} by up to an order of magnitude in both vertex-induced and edge-induced workloads. 
GraphMini has an overall better speedup for vertex-induced workloads because these patterns have a larger number of set operations to compute (due to anti-edges), providing more opportunities to accelerate set operations using auxiliary graphs. 

\begin{figure}
\begin{subfigure}{\linewidth}
  \centering
  \includegraphics[width=0.8 \linewidth]{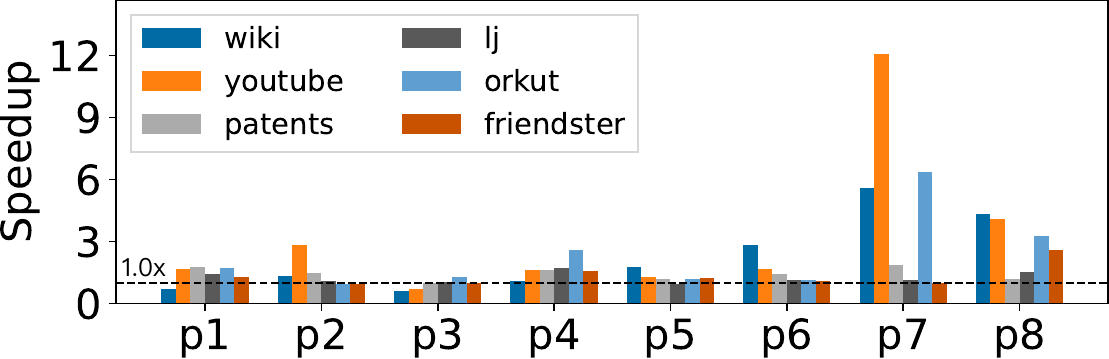}
  \caption{GraphMini vs Base (Edge-Induced)}
  \label{fig:mg_base_edge}
\end{subfigure} \newline
\begin{subfigure}{\linewidth}
  \centering
  \includegraphics[width=0.8 \linewidth]{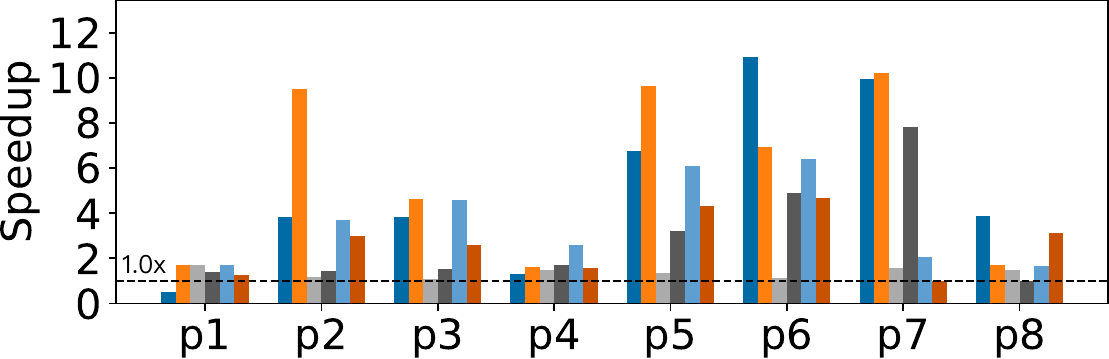}
  \caption{GraphMini vs Base (Vertex-Induced)}
  \label{fig:mg_base_vertex}
\end{subfigure} \newline
\caption{GraphMini vs Base}
\label{fig:mg_state_of_the_arts}
\end{figure}

Table~\ref{tab:comp_base} provides a breakdown of the performance enhancements using the methods introduced in this paper. $T_{Base}$ is the query execution time of \emph{Base}. 
$T_{Base+N}$ is the query execution time when we use nested loop parallelism on top of \emph{Base}. 
$T_{Online}$ considers using auxiliary graphs to accelerate set operations, but not nested loop parallelism. 
$T_{Online+N}$ considers both nested loop parallelism and auxiliary graphs, which is the default setting of GraphMini.

When the workload is initially imbalanced across different workers (ex. $P_6$ on wiki), the nested parallelism allows idle computation resources to participate in query execution, hence speeding up the query execution time. 
When the workload is more balanced (ex. $P_3$ and $P_4$ on Orkut), using auxiliary graphs accelerates set operations by reducing the total amount of work in absolute terms.
When the workload is imbalanced and set operations can be accelerated by using auxiliary graphs (ex. $P_5$ and $P_7$ on Wiki; $P_5$ on Orkut), combining both techniques dramatically speeds up query execution.

\begin{table}[]
\centering
\begin{tabular}{|c|ccc|ccc|}
\hline
\textbf{Graph}          & \multicolumn{3}{c|}{\textit{wiki}}                              & \multicolumn{3}{c|}{\textit{orkut}}                              \\ \hline
\textbf{Pattern} &
  \multicolumn{1}{c|}{\textit{$P_5$}} &
  \multicolumn{1}{c|}{\textit{$P_6$}} &
  \textit{$P_7$} &
  \multicolumn{1}{c|}{\textit{$P_3$}} &
  \multicolumn{1}{c|}{\textit{$P_4$}} &
  \textit{$P_5$} \\ \hline \hline
\textit{$T_{Base}$}     & \multicolumn{1}{c|}{6.4s}  & \multicolumn{1}{c|}{10.7s} & 33.8s & \multicolumn{1}{c|}{936s} & \multicolumn{1}{c|}{17.6s} & 35,790s \\ \hline
\textit{$T_{Online}$}   & \multicolumn{1}{c|}{2.6s}  & \multicolumn{1}{c|}{10.4s} & 15.8s & \multicolumn{1}{c|}{205s} & \multicolumn{1}{c|}{6.8s}  & 8,567s  \\ \hline
\textit{$T_{Base+N}$}   & \multicolumn{1}{c|}{2.5s}  & \multicolumn{1}{c|}{2.5s}  & 8.8s  & \multicolumn{1}{c|}{928s} & \multicolumn{1}{c|}{17.5s} & 30,542s \\ \hline
\textit{$T_{Online+N}$} & \multicolumn{1}{c|}{0.96s} & \multicolumn{1}{c|}{2.5s}  & 3.5s  & \multicolumn{1}{c|}{206s} & \multicolumn{1}{c|}{6.8s}  & 5,750s  \\ \hline \hline
\textbf{Speedup}       & \multicolumn{1}{c|}{6.6x}  & \multicolumn{1}{c|}{4.3x}  & 9.7x  & \multicolumn{1}{c|}{4.5x} & \multicolumn{1}{c|}{2.6x}  & 6.2x    \\ \hline
\end{tabular}
\caption{Comparison with Base (Vertex-Induced) (The speedup is computed via $T_{Base} / T_{Online + N}$). $T_{Online + N}$ corresponds to GraphMini. }
\label{tab:comp_base}
\end{table}

\begin{figure}[h]
    \centering
    \includegraphics[width=\linewidth]{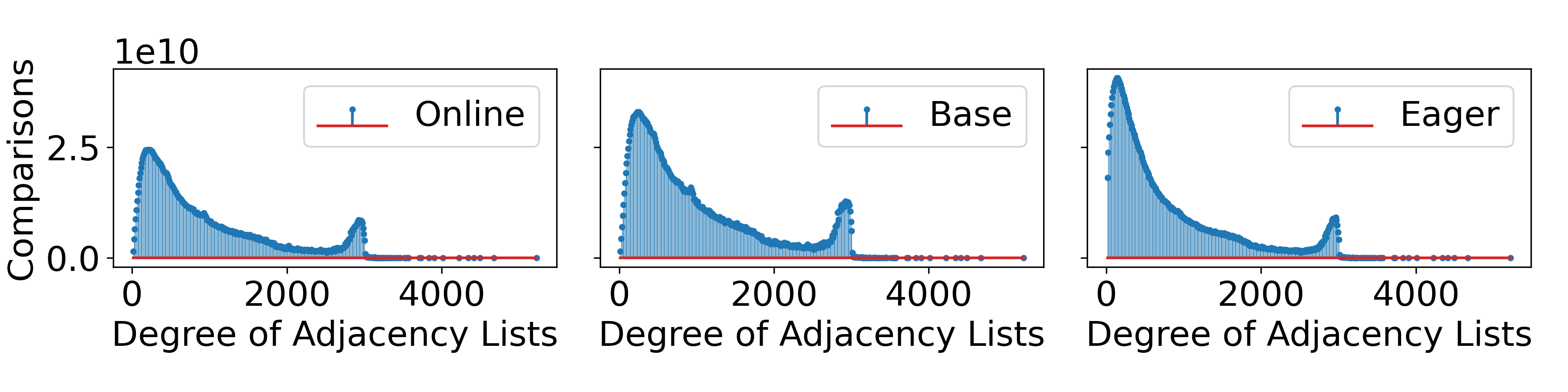}
    \caption{The total number of comparisons for running edge-induced $P_3$ on $Friendster$, aggregated by the length of the original adjacency lists. }
    \label{fig:p3_friendster}
\end{figure}
\spara{Online vs. Eager}
We now compare using GraphMini's online cost model against an \emph{Eager} variant that prunes all candidate adjacency lists and does not use the cost model.
\emph{Base} does not use auxiliary graphs at all.
We measure the number of comparisons used in the set operations they perform. 
\jl{For \emph{Online} and \emph{Eager}, the reported times include the set operations introduced to prune the adjacency lists.}
Figure~\ref{fig:p3_friendster} shows the total number of comparisons required to run $P_3$ (edge-induced) on $Friendster$, grouped by the degree of the original adjacency lists. 

The $Online$ approach has the overall least number of vertices scanned because it uses the cost model to avoid pruning adjacency lists that might not provide any benefit. 
The $Eager$ approach scans more vertices compared to the $Online$ and $Base$ to prune all candidate adjacency lists, but this cost is not amortized later during query execution.

\begin{table}[h]
\resizebox{\columnwidth}{!}{%
\begin{tabular}{|c|ccc|ccc|}
\hline
Pattern             & \multicolumn{3}{c|}{P4 (Vertex-Induced)}                                     & \multicolumn{3}{c|}{P5 (Vertex-Induced)}                                     \\ \hline
Graph               & \multicolumn{1}{c|}{$T_{w/o}$} & \multicolumn{1}{c|}{$T_{w}$}  & \textbf{Speedup} & \multicolumn{1}{c|}{$T_{w/o}$} & \multicolumn{1}{c|}{$T_{w}$}  & \textbf{Speedup} \\ \hline
\textit{Wiki}       & \multicolumn{1}{c|}{0.02s}  & \multicolumn{1}{c|}{0.02s} & 1.0x              & \multicolumn{1}{c|}{1.1s}   & \multicolumn{1}{c|}{0.9s}  & 1.2x              \\ \hline
\textit{Patents}    & \multicolumn{1}{c|}{0.08s}  & \multicolumn{1}{c|}{0.08s} & 1.0x              & \multicolumn{1}{c|}{0.4s}   & \multicolumn{1}{c|}{0.4s}  & 1.0x              \\ \hline
\textit{YouTube}    & \multicolumn{1}{c|}{0.03s}  & \multicolumn{1}{c|}{0.03s} & 1.0x              & \multicolumn{1}{c|}{4.6s}   & \multicolumn{1}{c|}{3.2s}  & 1.4x              \\ \hline
\textit{Lj}         & \multicolumn{1}{c|}{13.2s}  & \multicolumn{1}{c|}{13.2s} & 1.0x              & \multicolumn{1}{c|}{2651s}  & \multicolumn{1}{c|}{2523s} & 1.1x              \\ \hline
\textit{Orkut}      & \multicolumn{1}{c|}{8.6s}   & \multicolumn{1}{c|}{6.9s}  & 1.2x              & \multicolumn{1}{c|}{6831s}  & \multicolumn{1}{c|}{5861s} & 1.2x              \\ \hline
\textit{Friendster} & \multicolumn{1}{c|}{61.6s}  & \multicolumn{1}{c|}{50.1s} & 1.2x              & \multicolumn{1}{c|}{4558s}  & \multicolumn{1}{c|}{3385s} & 1.3x              \\ \hline
\end{tabular}%
}
\caption{Reusing auxiliary graphs for building auxiliary graphs.}
\label{tab:reuse}
\end{table}
\spara{Re-using Auxiliary Graphs}
Table~\ref{tab:reuse} shows the comparison of running $P_4$ and $P_5$ with and without reusing auxiliary graphs to build auxiliary graphs. For sparse graphs like $Wiki$, $Patents$, and $Lj$ the effect of reusing is not significant. On larger graphs like $Orkut$ and $Friendster$, reusing auxiliary graphs can further speed up query execution by $20\%$ - $30\%$. 

\begin{table}[]
\resizebox{\columnwidth}{!}{%
\begin{tabular}{|cc|cccccccc|}
\hline
                                 &       & \multicolumn{8}{c|}{Avg. size of auxiliary graphs (MB)}                                                                                                                                          \\ \hline
\multicolumn{2}{|c|}{Graph size (MB)}    & \multicolumn{1}{c|}{\textit{$P_1$}}  & \multicolumn{1}{c|}{\textit{$P_2$}}  & \multicolumn{1}{c|}{\textit{$P_3$}}  & \multicolumn{1}{c|}{\textit{$P_4$}}  & \multicolumn{1}{c|}{\textit{$P_5$}}  & \multicolumn{1}{c|}{\textit{$P_6$}}  & \multicolumn{1}{c|}{\textit{$P_7$}}  & \textit{$P_8$}  \\ \hline
\multicolumn{1}{|c|}{Wiki}       & 0.7   & \multicolumn{1}{c|}{0.1} & \multicolumn{1}{c|}{0.5} & \multicolumn{1}{c|}{0.4} & \multicolumn{1}{c|}{0.1} & \multicolumn{1}{c|}{0.5} & \multicolumn{1}{c|}{0.8} & \multicolumn{1}{c|}{0.6} & 0.3 \\ \hline
\multicolumn{1}{|c|}{Patents}    & 142   & \multicolumn{1}{c|}{0.1} & \multicolumn{1}{c|}{0.3} & \multicolumn{1}{c|}{0.2} & \multicolumn{1}{c|}{0.1} & \multicolumn{1}{c|}{0.3} & \multicolumn{1}{c|}{0.3} & \multicolumn{1}{c|}{0.7} & 0.3 \\ \hline
\multicolumn{1}{|c|}{YouTube}    & 28    & \multicolumn{1}{c|}{0.1} & \multicolumn{1}{c|}{10}  & \multicolumn{1}{c|}{2.0} & \multicolumn{1}{c|}{0.1} & \multicolumn{1}{c|}{2.3} & \multicolumn{1}{c|}{3.7} & \multicolumn{1}{c|}{80}  & 0.4 \\ \hline
\multicolumn{1}{|c|}{Lj}         & 281   & \multicolumn{1}{c|}{1.2} & \multicolumn{1}{c|}{7.1} & \multicolumn{1}{c|}{6.6} & \multicolumn{1}{c|}{2.1} & \multicolumn{1}{c|}{6.8} & \multicolumn{1}{c|}{5.5} & \multicolumn{1}{c|}{7.3} & OT  \\ \hline
\multicolumn{1}{|c|}{Orkut}      & 907   & \multicolumn{1}{c|}{0.9} & \multicolumn{1}{c|}{164} & \multicolumn{1}{c|}{56}  & \multicolumn{1}{c|}{1.0} & \multicolumn{1}{c|}{59}  & \multicolumn{1}{c|}{59}  & \multicolumn{1}{c|}{113} & 9.5 \\ \hline
\multicolumn{1}{|c|}{Friendster} & 14.4k & \multicolumn{1}{c|}{0.6} & \multicolumn{1}{c|}{39}  & \multicolumn{1}{c|}{37}  & \multicolumn{1}{c|}{0.7} & \multicolumn{1}{c|}{37}  & \multicolumn{1}{c|}{20}  & \multicolumn{1}{c|}{OT}  & 3.1 \\ \hline
\end{tabular}%
}
\caption{Average memory consumption for storing auxiliary graphs per thread. (OT means the query takes more than 24 hours to run)}
\label{tab:memory}
\end{table}
\subsection{Memory Consumption}
We now discuss the memory consumption of storing auxiliary graphs during query execution. 
Table~\ref{tab:memory} shows the average memory requirement per thread for storing auxiliary graphs in the vertex-induced variant. 
The memory footprint of the edge-induced variant is smaller than the vertex-induced one because there are no anti-edges, so we don't need to build auxiliary graphs to accelerate set subtractions.

The results show that the memory consumption for storing auxiliary graphs on each worker thread is usually only a small fraction of the data graph. 
Memory usage for storing auxiliary graphs is associated with three main factors: the maximum vertex degree in the data graph, the density of the query graph, and the canonicality constraints of the query schedule. 

The maximum vertex degree in the data graph impacts memory consumption because it determines the maximum size of the prefix sets, which are used to build auxiliary graphs. For example, $Wiki$ and $Patents$ have similar maximum degrees, and the memory consumption of auxiliary graphs is similar on these graphs. The maximum degree in \emph{Orkut} is larger than that of \emph{Friendster} so the memory consumption on Orkut is larger across all queries, even though Friendster is a much larger data graph than Orkut. 

The density of the query graph impacts memory consumption because denser patterns (e.g., cliques) perform more set intersections and present more opportunities to prune adjacency lists.
In vertex-induced matching, the absence of an edge between two query vertices implies the presence of an anti-edge, which enables the use of set subtractions to compute the prefix sets.
However, the pruning power of set subtraction is usually less than that of set intersections on sparse data graphs.
As a result, the prefix sets will have a larger size which leads to larger memory footprints when building auxiliary graphs.
For example, the query graph $P_2$ is less dense than $P_3$ and $P_4$ and therefore it has the largest memory consumption. 

The canonicality constraints of the query schedule also play an important role because they enhance the pruning power of set intersections.
This power depends on multiple factors such as the symmetry of the query graph~\cite{GraphPi,GraphZero}.
For example, the schedules for query $P_6$ have stricter canonicality constraints than those of $P_5$, so in some cases auxiliary graphs for $P_6$ have a lower memory consumption than for $P_5$ even though $P_5$ is denser than $P_6$.

\section{Related work}
Several frameworks \cite{Pregel, GraphX, Ligra} tackle general graph processing. 
These frameworks provide an extensible high-level task API while abstracting several low-level components such as scheduling, graph access, and parallelism. However, they are not well suited for graph mining problems such as motif counting and frequent subgraph mining. 
This has motivated the development of dedicated subgraph-centric graph mining frameworks \cite{Arabesque, RStream, Pangolin, Fractal, Peregrine, AutoMine, GraphZero, GraphPi}. 
Recent graph mining systems work uses subgraph query matching as the key building block \cite{AutoMine, Peregrine}. 
They select the optimal schedules at query time and generate an optimized algorithm or binary to handle query matching. 
State-of-the-art systems use a code generation approach to improve performance further \cite{ AutoMine, GraphZero}. 
GraphMini advances this line of work by proposing a novel direction for performance optimization.
Throughout the paper, we extensively discuss the state-of-the-art to motivate our work and show that GraphMini outperforms it. 

This work considers \emph{subgraph enumeration} using unlabeled data graphs and queries.
Other prior work focuses on \emph{subgraph matching}, where graphs and queries are labeled.
For a recent survey and experimental evaluation of these algorithms, see~\cite{in-depth}.
RapidMatch is another more recent work in this area~\cite{rapidmatch}.
cuTS optimizes subgraph matching on GPUs~\cite{cuTS}. 
A key challenge in subgraph matching is leveraging labels to filter candidate vertices and find a good query schedule, which is not an issue in subgraph enumeration.
Auxiliary data structures are commonly used in subgraph matching \cite{GraphQL,BI,Han,in-depth,CECI,Turboiso,Versatile} and lead to significant speedups \cite{in-depth, Versatile}. 
However, these techniques rely on labels and are not designed for subgraph enumeration. 

This work considers main-memory graphs.
Other work considered algorithms for out-of-core graphs~\cite{DUALSIM}, or large graphs that are partitioned across servers~\cite{yang2021huge}.
Orthogonal optimizations include exploiting symmetry to reuse computation~\cite{ParallelSingleMachine} or exploring output compression~\cite{cbf}. 
PBE-REUSE shares the results of set intersections across GPU threads by using a BFS matching strategy, which results in a much larger intermediate state than the more common backtracking DFS approach~\cite{guo2020exploiting}. 
None of these algorithms considered proactive online graph pruning,  which is introduced in this work.

Some work proposes accelerating set intersections by using a graph's binary representation and AVX/SSE instructions in the CPUs \cite{EmptyHeaded, SIMD}. 
Some work proposes using specialized hardware to accelerate subgraph matching. FlexMiner \cite{FlexMiner} uses specialized hardware and on-chip memory to accelerate subgraph enumeration. DIMMining  \cite{DIMMining} is a more recent work that uses newly designed instruction sets and hardware to accelerate set operations in subgraph enumeration. Our work also aims to speed up set operations, but it does so by reducing the size of the input sets, independent of how the inputs are represented. These two approaches are complementary, and combining them is an interesting direction. 

Graph databases are increasingly used in many applications and have proposed several techniques to optimize the query plan~\cite{GraphProcess,Demystify,EmptyHeaded,JoinProcessing,mhedhbi2019optimizing}.
Query optimization is an orthogonal problem in our work.

\section{Conclusion}
Graph pattern matching is a fundamental problem.
Auxiliary graphs represent a novel direction to reduce the running time of graph pattern mining algorithms.
It entails building a pruned representation of the data graph that can be used instead of the original graph to speed up query execution.
Auxiliary graphs are complementary to and can be combined with other state-of-the-art algorithms for query scheduling and code generation.
The GraphMini system shows that auxiliary graphs can speed up query execution time by up to $10\times$ with a low memory cost.
Generating query execution code with auxiliary graphs introduces a negligible compilation overhead.
Proactive graph pruning is a promising avenue to speed up subgraph enumeration and potentially other graph pattern-matching problems.

\section*{Acknowledgements}

We want to thank Alexandra Meliou, Peter Haas, our shepherd Ana-Lucia Varbanescu, and the anonymous reviewers for their insightful feedback.
This work was supported by the National Science Foundation under Grant No. CNS-2224054.
Any opinions, findings, conclusions, and recommendations expressed in this material are those of the author(s) and do not necessarily reflect the views of the National Science Foundation.
The work was also supported in part by an Amazon Research Award and an Adobe Research Collaboration Grant.

\section*{Artifacts}
\subsection{Artifact and experiment requirements}

Our artifact is publicly available at \url{https://zenodo.org/record/8350615}. The experiment compares the query execution and compilation time of the baseline systems with those of GraphMini.

\spara{Hardware} We run all the experiments on an AWS EC2 c6i.16xlarge instance equipped with a 32-core Intel Xeon 8375C CPU and 128GB host memory.

\spara{Software} We build all tested systems using GCC (v11.3) with C++17 standard and -O3 optimization flag. We use CMake (v3.20) as the build system. We use the original code base of GraphPi and Dryadic for comparison, except for minor changes in data loading and result output. 

\spara{Dataset} We use six publicly available datasets in our experiments: Wiki, YouTube, Patents, LiveJournal, Orkut, and Friendster. All the datasets are downloaded from \url{https://snap.stanford.edu/data/index.html}.

\subsection{Procedures to reproduce the experiments}
We provide scripts for installing dependent libraries, downloading datasets, and pre-processing the datasets into the required format. See the document in the artifacts for more details. 
We provide scripts to reproduce the benchmark experiments, which contain running scripts to use Dryadic, GraphPi, and GraphMini on different data graphs and query patterns. 
We provide a document (README.md) in the artifact to explain how to use these scripts to automate the experiment reproduction process. 

\bibliographystyle{IEEEtran}
\bibliography{Minigraph}

\begin{thebibliography}{10}
\providecommand{\url}[1]{#1}
\csname url@samestyle\endcsname
\providecommand{\newblock}{\relax}
\providecommand{\bibinfo}[2]{#2}
\providecommand{\BIBentrySTDinterwordspacing}{\spaceskip=0pt\relax}
\providecommand{\BIBentryALTinterwordstretchfactor}{4}
\providecommand{\BIBentryALTinterwordspacing}{\spaceskip=\fontdimen2\font plus
\BIBentryALTinterwordstretchfactor\fontdimen3\font minus
  \fontdimen4\font\relax}
\providecommand{\BIBforeignlanguage}[2]{{%
\expandafter\ifx\csname l@#1\endcsname\relax
\typeout{** WARNING: IEEEtran.bst: No hyphenation pattern has been}%
\typeout{** loaded for the language `#1'. Using the pattern for}%
\typeout{** the default language instead.}%
\else
\language=\csname l@#1\endcsname
\fi
#2}}
\providecommand{\BIBdecl}{\relax}
\BIBdecl

\bibitem{social1}
W.~Fan, ``Graph pattern matching revised for social network analysis,''
  \emph{ACM International Conference Proceeding Series}, 03 2012.

\bibitem{social2}
\BIBentryALTinterwordspacing
T.~A.~B. Snijders, P.~E. Pattison, G.~L. Robins, and M.~S. Handcock, ``New
  specifications for exponential random graph models,'' \emph{Sociological
  Methodology}, vol.~36, no.~1, pp. 99--153, 2006. [Online]. Available:
  \url{https://doi.org/10.1111/j.1467-9531.2006.00176.x}
\BIBentrySTDinterwordspacing

\bibitem{ppt1}
N.~Alon, P.~Dao, I.~Hajirasouliha, F.~Hormozdiari, and C.~Sahinalp,
  ``Biomolecular network motif counting and discovery by color coding,''
  \emph{Bioinformatics (Oxford, England)}, vol.~24, pp. i241--9, 07 2008.

\bibitem{Peregrine}
\BIBentryALTinterwordspacing
K.~Jamshidi, R.~Mahadasa, and K.~Vora, ``Peregrine: A pattern-aware graph
  mining system,'' in \emph{Proceedings of the Fifteenth European Conference on
  Computer Systems}, ser. EuroSys '20.\hskip 1em plus 0.5em minus 0.4em\relax
  New York, NY, USA: Association for Computing Machinery, 2020. [Online].
  Available: \url{https://doi.org/10.1145/3342195.3387548}
\BIBentrySTDinterwordspacing

\bibitem{AutoMine}
\BIBentryALTinterwordspacing
D.~Mawhirter and B.~Wu, ``Automine: Harmonizing high-level abstraction and high
  performance for graph mining,'' in \emph{Proceedings of the 27th ACM
  Symposium on Operating Systems Principles}, ser. SOSP '19.\hskip 1em plus
  0.5em minus 0.4em\relax New York, NY, USA: Association for Computing
  Machinery, 2019, p. 509–523. [Online]. Available:
  \url{https://doi.org/10.1145/3341301.3359633}
\BIBentrySTDinterwordspacing

\bibitem{GraphZero}
\BIBentryALTinterwordspacing
D.~Mawhirter, S.~Reinehr, C.~Holmes, T.~Liu, and B.~Wu, ``Graphzero: Breaking
  symmetry for efficient graph mining,'' 2019. [Online]. Available:
  \url{https://arxiv.org/abs/1911.12877}
\BIBentrySTDinterwordspacing

\bibitem{Dryadic}
\BIBentryALTinterwordspacing
D.~Mawhirter, S.~Reinehr, W.~Han, N.~Fields, M.~Claver, C.~Holmes, J.~McClurg,
  T.~Liu, and B.~Wu, ``Dryadic: Flexible and fast graph pattern matching at
  scale,'' in \emph{30th International Conference on Parallel Architectures and
  Compilation Techniques, {PACT} 2021, Atlanta, GA, USA, September 26-29,
  2021}, J.~Lee and A.~Cohen, Eds.\hskip 1em plus 0.5em minus 0.4em\relax
  {IEEE}, 2021, pp. 289--303. [Online]. Available:
  \url{https://doi.org/10.1109/PACT52795.2021.00028}
\BIBentrySTDinterwordspacing

\bibitem{SIMD}
\BIBentryALTinterwordspacing
S.~Han, L.~Zou, and J.~X. Yu, ``Speeding up set intersections in graph
  algorithms using simd instructions,'' in \emph{Proceedings of the 2018
  International Conference on Management of Data}, ser. SIGMOD '18.\hskip 1em
  plus 0.5em minus 0.4em\relax New York, NY, USA: Association for Computing
  Machinery, 2018, p. 1587–1602. [Online]. Available:
  \url{https://doi.org/10.1145/3183713.3196924}
\BIBentrySTDinterwordspacing

\bibitem{GraphPi}
T.~Shi, M.~Zhai, Y.~Xu, and J.~Zhai, ``Graphpi: High performance graph pattern
  matching through effective redundancy elimination,'' in \emph{SC20:
  International Conference for High Performance Computing, Networking, Storage
  and Analysis}, 2020, pp. 1--14.

\bibitem{EmptyHeaded}
\BIBentryALTinterwordspacing
C.~R. Aberger, A.~Lamb, S.~Tu, A.~N\"{o}tzli, K.~Olukotun, and C.~R\'{e},
  ``Emptyheaded: A relational engine for graph processing,'' \emph{ACM Trans.
  Database Syst.}, vol.~42, no.~4, oct 2017. [Online]. Available:
  \url{https://doi.org/10.1145/3129246}
\BIBentrySTDinterwordspacing

\bibitem{FlexMiner}
X.~Chen, T.~Huang, S.~Xu, T.~Bourgeat, C.~Chung, and A.~Arvind, ``Flexminer: A
  pattern-aware accelerator for graph pattern mining,'' in \emph{2021 ACM/IEEE
  48th Annual International Symposium on Computer Architecture (ISCA)}, 2021,
  pp. 581--594.

\bibitem{DIMMining}
\BIBentryALTinterwordspacing
G.~Dai, Z.~Zhu, T.~Fu, C.~Wei, B.~Wang, X.~Li, Y.~Xie, H.~Yang, and Y.~Wang,
  ``Dimmining: Pruning-efficient and parallel graph mining on
  near-memory-computing,'' in \emph{Proceedings of the 49th Annual
  International Symposium on Computer Architecture}, ser. ISCA '22.\hskip 1em
  plus 0.5em minus 0.4em\relax New York, NY, USA: Association for Computing
  Machinery, 2022, p. 130–145. [Online]. Available:
  \url{https://doi.org/10.1145/3470496.3527388}
\BIBentrySTDinterwordspacing

\bibitem{Arabesque}
\BIBentryALTinterwordspacing
C.~H.~C. Teixeira, A.~J. Fonseca, M.~Serafini, G.~Siganos, M.~J. Zaki, and
  A.~Aboulnaga, ``Arabesque: A system for distributed graph mining,'' in
  \emph{Proceedings of the 25th Symposium on Operating Systems Principles},
  ser. SOSP '15.\hskip 1em plus 0.5em minus 0.4em\relax New York, NY, USA:
  Association for Computing Machinery, 2015, p. 425–440. [Online]. Available:
  \url{https://doi.org/10.1145/2815400.2815410}
\BIBentrySTDinterwordspacing

\bibitem{Ullmann}
\BIBentryALTinterwordspacing
J.~R. Ullmann, ``An algorithm for subgraph isomorphism,'' \emph{J. ACM},
  vol.~23, no.~1, p. 31–42, jan 1976. [Online]. Available:
  \url{https://doi.org/10.1145/321921.321925}
\BIBentrySTDinterwordspacing

\bibitem{SNAP_datasets}
J.~Leskovec and A.~Krevl, ``{SNAP Datasets}: {Stanford} large network dataset
  collection,'' \url{http://snap.stanford.edu/data}, Jun. 2014.

\bibitem{Pregel}
\BIBentryALTinterwordspacing
G.~Malewicz, M.~H. Austern, A.~J. Bik, J.~C. Dehnert, I.~Horn, N.~Leiser, and
  G.~Czajkowski, ``Pregel: A system for large-scale graph processing,'' in
  \emph{Proceedings of the 2010 ACM SIGMOD International Conference on
  Management of Data}, ser. SIGMOD '10.\hskip 1em plus 0.5em minus 0.4em\relax
  New York, NY, USA: Association for Computing Machinery, 2010, p. 135–146.
  [Online]. Available: \url{https://doi.org/10.1145/1807167.1807184}
\BIBentrySTDinterwordspacing

\bibitem{GraphX}
J.~E. Gonzalez, R.~S. Xin, A.~Dave, D.~Crankshaw, M.~J. Franklin, and
  I.~Stoica, ``Graphx: Graph processing in a distributed dataflow framework,''
  in \emph{Proceedings of the 11th USENIX Conference on Operating Systems
  Design and Implementation}, ser. OSDI'14.\hskip 1em plus 0.5em minus
  0.4em\relax USA: USENIX Association, 2014, p. 599–613.

\bibitem{Ligra}
\BIBentryALTinterwordspacing
J.~Shun and G.~E. Blelloch, ``Ligra: A lightweight graph processing framework
  for shared memory,'' in \emph{Proceedings of the 18th ACM SIGPLAN Symposium
  on Principles and Practice of Parallel Programming}, ser. PPoPP '13.\hskip
  1em plus 0.5em minus 0.4em\relax New York, NY, USA: Association for Computing
  Machinery, 2013, p. 135–146. [Online]. Available:
  \url{https://doi.org/10.1145/2442516.2442530}
\BIBentrySTDinterwordspacing

\bibitem{RStream}
K.~Wang, Z.~Zuo, J.~Thorpe, T.~Q. Nguyen, and G.~H. Xu, ``Rstream: Marrying
  relational algebra with streaming for efficient graph mining on a single
  machine,'' in \emph{Proceedings of the 13th USENIX Conference on Operating
  Systems Design and Implementation}, ser. OSDI'18.\hskip 1em plus 0.5em minus
  0.4em\relax USA: USENIX Association, 2018, p. 763–782.

\bibitem{Pangolin}
\BIBentryALTinterwordspacing
X.~Chen, R.~Dathathri, G.~Gill, and K.~Pingali, ``Pangolin: An efficient and
  flexible graph mining system on cpu and gpu,'' \emph{Proc. VLDB Endow.},
  vol.~13, no.~8, p. 1190–1205, apr 2020. [Online]. Available:
  \url{https://doi.org/10.14778/3389133.3389137}
\BIBentrySTDinterwordspacing

\bibitem{Fractal}
\BIBentryALTinterwordspacing
V.~Dias, C.~H.~C. Teixeira, D.~Guedes, W.~Meira, and S.~Parthasarathy,
  ``Fractal: A general-purpose graph pattern mining system,'' in
  \emph{Proceedings of the 2019 International Conference on Management of
  Data}, ser. SIGMOD '19.\hskip 1em plus 0.5em minus 0.4em\relax New York, NY,
  USA: Association for Computing Machinery, 2019, p. 1357–1374. [Online].
  Available: \url{https://doi.org/10.1145/3299869.3319875}
\BIBentrySTDinterwordspacing

\bibitem{in-depth}
\BIBentryALTinterwordspacing
S.~Sun and Q.~Luo, ``In-memory subgraph matching: An in-depth study,'' in
  \emph{Proceedings of the 2020 ACM SIGMOD International Conference on
  Management of Data}, ser. SIGMOD '20.\hskip 1em plus 0.5em minus 0.4em\relax
  New York, NY, USA: Association for Computing Machinery, 2020, p. 1083–1098.
  [Online]. Available: \url{https://doi.org/10.1145/3318464.3380581}
\BIBentrySTDinterwordspacing

\bibitem{rapidmatch}
\BIBentryALTinterwordspacing
S.~Sun, X.~Sun, Y.~Che, Q.~Luo, and B.~He, ``Rapidmatch: A holistic approach to
  subgraph query processing,'' \emph{Proc. VLDB Endow.}, vol.~14, no.~2, p.
  176–188, oct 2020. [Online]. Available:
  \url{https://doi.org/10.14778/3425879.3425888}
\BIBentrySTDinterwordspacing

\bibitem{cuTS}
\BIBentryALTinterwordspacing
L.~Xiang, A.~Khan, E.~Serra, M.~Halappanavar, and A.~Sukumaran-Rajam, ``Cuts:
  Scaling subgraph isomorphism on distributed multi-gpu systems using trie
  based data structure,'' in \emph{Proceedings of the International Conference
  for High Performance Computing, Networking, Storage and Analysis}, ser. SC
  '21.\hskip 1em plus 0.5em minus 0.4em\relax New York, NY, USA: Association
  for Computing Machinery, 2021. [Online]. Available:
  \url{https://doi.org/10.1145/3458817.3476214}
\BIBentrySTDinterwordspacing

\bibitem{GraphQL}
\BIBentryALTinterwordspacing
H.~He and A.~K. Singh, ``Graphs-at-a-time: Query language and access methods
  for graph databases,'' in \emph{Proceedings of the 2008 ACM SIGMOD
  International Conference on Management of Data}, ser. SIGMOD '08.\hskip 1em
  plus 0.5em minus 0.4em\relax New York, NY, USA: Association for Computing
  Machinery, 2008, p. 405–418. [Online]. Available:
  \url{https://doi.org/10.1145/1376616.1376660}
\BIBentrySTDinterwordspacing

\bibitem{BI}
\BIBentryALTinterwordspacing
F.~Bi, L.~Chang, X.~Lin, L.~Qin, and W.~Zhang, ``Efficient subgraph matching by
  postponing cartesian products,'' in \emph{Proceedings of the 2016
  International Conference on Management of Data}, ser. SIGMOD '16.\hskip 1em
  plus 0.5em minus 0.4em\relax New York, NY, USA: Association for Computing
  Machinery, 2016, p. 1199–1214. [Online]. Available:
  \url{https://doi.org/10.1145/2882903.2915236}
\BIBentrySTDinterwordspacing

\bibitem{Han}
\BIBentryALTinterwordspacing
M.~Han, H.~Kim, G.~Gu, K.~Park, and W.-S. Han, ``Efficient subgraph matching:
  Harmonizing dynamic programming, adaptive matching order, and failing set
  together,'' in \emph{Proceedings of the 2019 International Conference on
  Management of Data}, ser. SIGMOD '19.\hskip 1em plus 0.5em minus 0.4em\relax
  New York, NY, USA: Association for Computing Machinery, 2019, p. 1429–1446.
  [Online]. Available: \url{https://doi.org/10.1145/3299869.3319880}
\BIBentrySTDinterwordspacing

\bibitem{CECI}
\BIBentryALTinterwordspacing
B.~Bhattarai, H.~Liu, and H.~H. Huang, ``Ceci: Compact embedding cluster index
  for scalable subgraph matching,'' in \emph{Proceedings of the 2019
  International Conference on Management of Data}, ser. SIGMOD '19.\hskip 1em
  plus 0.5em minus 0.4em\relax New York, NY, USA: Association for Computing
  Machinery, 2019, p. 1447–1462. [Online]. Available:
  \url{https://doi.org/10.1145/3299869.3300086}
\BIBentrySTDinterwordspacing

\bibitem{Turboiso}
\BIBentryALTinterwordspacing
W.-S. Han, J.~Lee, and J.-H. Lee, ``Turboiso: Towards ultrafast and robust
  subgraph isomorphism search in large graph databases,'' in \emph{Proceedings
  of the 2013 ACM SIGMOD International Conference on Management of Data}, ser.
  SIGMOD '13.\hskip 1em plus 0.5em minus 0.4em\relax New York, NY, USA:
  Association for Computing Machinery, 2013, p. 337–348. [Online]. Available:
  \url{https://doi.org/10.1145/2463676.2465300}
\BIBentrySTDinterwordspacing

\bibitem{Versatile}
\BIBentryALTinterwordspacing
H.~Kim, Y.~Choi, K.~Park, X.~Lin, S.-H. Hong, and W.-S. Han, ``Versatile
  equivalences: Speeding up subgraph query processing and subgraph matching,''
  in \emph{Proceedings of the 2021 International Conference on Management of
  Data}, ser. SIGMOD '21.\hskip 1em plus 0.5em minus 0.4em\relax New York, NY,
  USA: Association for Computing Machinery, 2021, p. 925–937. [Online].
  Available: \url{https://doi.org/10.1145/3448016.3457265}
\BIBentrySTDinterwordspacing

\bibitem{DUALSIM}
\BIBentryALTinterwordspacing
H.~Kim, J.~Lee, S.~S. Bhowmick, W.-S. Han, J.~Lee, S.~Ko, and M.~H. Jarrah,
  ``Dualsim: Parallel subgraph enumeration in a massive graph on a single
  machine,'' in \emph{Proceedings of the 2016 International Conference on
  Management of Data}, ser. SIGMOD '16.\hskip 1em plus 0.5em minus 0.4em\relax
  New York, NY, USA: Association for Computing Machinery, 2016, p. 1231–1245.
  [Online]. Available: \url{https://doi.org/10.1145/2882903.2915209}
\BIBentrySTDinterwordspacing

\bibitem{yang2021huge}
Z.~Yang, L.~Lai, X.~Lin, K.~Hao, and W.~Zhang, ``Huge: An efficient and
  scalable subgraph enumeration system,'' in \emph{Proceedings of the 2021
  International Conference on Management of Data}, 2021, pp. 2049--2062.

\bibitem{ParallelSingleMachine}
S.~Sun, Y.~Che, L.~Wang, and Q.~Luo, ``Efficient parallel subgraph enumeration
  on a single machine,'' in \emph{2019 IEEE 35th International Conference on
  Data Engineering (ICDE)}, 2019, pp. 232--243.

\bibitem{cbf}
\BIBentryALTinterwordspacing
M.~Qiao, H.~Zhang, and H.~Cheng, ``Subgraph matching: On compression and
  computation,'' \emph{Proc. VLDB Endow.}, vol.~11, no.~2, p. 176–188, oct
  2017. [Online]. Available: \url{https://doi.org/10.14778/3149193.3149198}
\BIBentrySTDinterwordspacing

\bibitem{guo2020exploiting}
W.~Guo, Y.~Li, and K.-L. Tan, ``Exploiting reuse for gpu subgraph
  enumeration,'' \emph{IEEE Transactions on Knowledge and Data Engineering},
  2020.

\bibitem{GraphProcess}
\BIBentryALTinterwordspacing
S.~Sahu, A.~Mhedhbi, S.~Salihoglu, J.~Lin, and M.~T. \"{O}zsu, ``The ubiquity
  of large graphs and surprising challenges of graph processing,'' \emph{Proc.
  VLDB Endow.}, vol.~11, no.~4, p. 420–431, dec 2017. [Online]. Available:
  \url{https://doi.org/10.1145/3164135.3164139}
\BIBentrySTDinterwordspacing

\bibitem{Demystify}
\BIBentryALTinterwordspacing
M.~Besta, E.~Peter, R.~Gerstenberger, M.~Fischer, M.~Podstawski, C.~Barthels,
  G.~Alonso, and T.~Hoefler, ``Demystifying graph databases: Analysis and
  taxonomy of data organization, system designs, and graph queries,'' 2019.
  [Online]. Available: \url{https://arxiv.org/abs/1910.09017}
\BIBentrySTDinterwordspacing

\bibitem{JoinProcessing}
\BIBentryALTinterwordspacing
D.~Nguyen, M.~Aref, M.~Bravenboer, G.~Kollias, H.~Q. Ngo, C.~R\'{e}, and
  A.~Rudra, ``Join processing for graph patterns: An old dog with new tricks,''
  in \emph{Proceedings of the GRADES'15}, ser. GRADES'15.\hskip 1em plus 0.5em
  minus 0.4em\relax New York, NY, USA: Association for Computing Machinery,
  2015. [Online]. Available: \url{https://doi.org/10.1145/2764947.2764948}
\BIBentrySTDinterwordspacing

\bibitem{mhedhbi2019optimizing}
A.~Mhedhbi and S.~Salihoglu, ``Optimizing subgraph queries by combining binary
  and worst-case optimal joins,'' \emph{arXiv preprint arXiv:1903.02076}, 2019.

\end{thebibliography}
\end{document}